\documentclass[aps,pra,twocolumn,showpacs,groupedaddress,floatfix]{revtex4}
\usepackage{amsfonts}
\usepackage{amsmath}
\usepackage{amssymb}
\usepackage{epsfig}
\usepackage{graphicx}
\begin{document}
\title{Improving the entanglement transfer from continuous
variable systems to localized qubits using non Gaussian states}
\date{\today}
\author{Federico Casagrande}
\email{federico.casagrande@mi.infn.it}
\author{Alfredo Lulli}
\email{alfredo.lulli@unimi.it}
\author{Matteo G.A. Paris}
\email{matteo.paris@fisica.unimi.it}
\affiliation{Dipartimento di Fisica dell'Universit\`{a} di
Milano, Italia}
\begin{abstract}
We investigate the entanglement transfer from a bipartite continuous-variable
(CV) system to a pair of localized qubits assuming that each CV mode couples
to one qubit via the off-resonance Jaynes-Cummings interaction with different
interaction times for the two subsystems. First, we consider the case of the
CV system prepared in a Bell-like superposition and investigate the conditions
for maximum entanglement transfer. Then we analyze the general case of two-mode
CV states that can be represented by a Schmidt decomposition in the Fock
number basis. This class includes both Gaussian and non Gaussian CV states, as
for example twin-beam (TWB) and pair-coherent (TMC, also known as two-mode-coherent)
states respectively. Under resonance conditions, equal interaction times for both
qubits and different initial preparations, we find that the entanglement transfer
is more efficient for TMC than for TWB states. In the perspective of
applications such as in cavity QED or with superconducting qubits, we
analyze in details the effects of off-resonance interactions (detuning) and different
interaction times for the two qubits, and discuss conditions to preserve the
entanglement transfer.
\end{abstract}
\maketitle
\section{Introduction}
Entanglement is the main resource of quantum information processing
(QIP). Indeed, much attention has been devoted to generation and
manipulation of entanglement either in discrete or in continuous
variable (CV) systems. Crucial and rewarding steps in the
development of QIP are now the storage of entanglement in quantum
memories~\cite{mem,mem1} and the transfer of entanglement from
localized to flying registers and viceversa. Indeed, effective
protocols for the distribution of entanglement would allow one to
realize quantum cryptography over long distances~\cite{qcr}, as well
as distributed quantum computation~\cite{dqc} and distributed
network for quantum communication purposes.\\
\indent Few schemes have been suggested either to entangle localized
qubits, e.g. distant atoms or superconducting quantum interference
devices, using squeezed radiation~\cite{Kraus Cirac 2004} or to
transfer entanglement between qubits and radiation~\cite{Paternostro
2004,Paternostro2 2004,Paternostro3 2004}. As a matter of fact,
efficient sources of entanglement have been developed for CV
systems, especially by quantum-optical implementations~\cite{CV
Fields}.  Indeed, multiphoton states might be optimal when
considering long distance communication, where they may travel
through free space or optical fibers, in view of the robustness of
their entanglement against losses ~\cite{cv-ent}.\\
\indent The entanglement transfer from free propagating light to
atomic systems has been achieved experimentally in the recent
years~\cite{mem,Hald 2000}. From the theoretical point of view,
the resonant entanglement transfer between a bipartite continuous
variable systems and a pair of qubits has been firstly analyzed
in~\cite{Son 2002} where the CV field is assumed to be a two-mode
squeezed vacuum or twin-beam (TWB) state~\cite{TWB} with the two
modes injected into spatially separate cavities. Two identical
atoms, both in the ground state, are then assumed to interact
resonantly, one for each cavity, with the cavity mode field for an
interaction time shorter than the cavity lifetime.  More recently,
a general approach has been developed~\cite{Paternostro 2004}, in
which two static qubits are isolated by the real world by their
own single mode bosonic local environment that also rules the
interaction of each qubit with an external driving field assumed
to be a general broadband two mode field. This model may be
applied to describe  a cavity QED setup with two atomic qubits
trapped into remote cavities. In Ref.~\cite{Paternostro2 2004} the
problems related to different interaction times for the two qubits
are pointed out, either for atomic qubits or in the case of
superconducting quantum interference devices (SQUID) qubits. The
possibility to transfer the entanglement of a TWB radiation field
to SQUIDS has been also
investigated in~\cite{Paternostro3 2004}.\\
\indent Very recently, in~\cite{Zou 2006} the entanglement transfer
process between CV and qubit bipartite systems was investigated.
Their scheme is composed by two atoms placed into two spatially
separated identical cavities where the two modes are injected. They
consider resonant interaction of two-mode fields, such as two-photon
superpositions, entangled coherent states and TWB, discussing
conditions for maximum entanglement transfer.\\
\indent The inverse problem of entanglement reciprocation from
qubits to continuous variables has been discussed in~\cite{Lee 2006}
by means of a model involving two atoms prepared in a maximally
entangled state and then injected into two spatially separated
cavities each one prepared in a coherent state. It was shown that
when the atoms leave the cavity their entanglement is transferred to
the post selected cavity fields.  The generated field entanglement
can be then transferred back to qubits, {\em i.e} to another couple
of atoms flying through the cavities. In a recent paper~\cite{McHugh
2006} the relationship between entanglement, mixedness and energy of
two qubits and two mode Gaussian quantum states has been analyzed,
whereas a strategy to enhance the entanglement transfer between TWB
states and multiple qubits has been suggested in~\cite{Serafini
2006}.\\
\indent In this paper we investigate the dynamics of a two-mode
entangled state of radiation coupled to a pair of localized qubits
via the off-resonance Jaynes-Cummings interaction. We focus our
attention on the entanglement transfer from radiation to atomic
qubits, though our analysis may be employed also to describe the
effective interaction of radiation with superconducting qubits. In
particular, compared to previous analysis, we consider in details
the effects of off-resonance interactions (detuning) and different
interaction times for the two qubits.  As a carrier of entanglement
we consider the general case of two-mode states that can be
represented by a Schmidt decomposition in the Fock number basis.
These include Gaussian states of radiation like twin-beams, realized
by nondegenerate parametric amplifiers by means of spontaneous
downconversion in nonlinear crystals, as well as non Gaussian
states, as for example pair-coherent (TMC, also known as two-mode
coherent) states~\cite{PC}, that can be obtained either by
degenerate Raman processes \cite{Zheng} or, more realistically, by
conditional measurements \cite{zou} and nondegenerate parametric
oscillators~\cite{PC2,PC3}. In fact, we find that TMC are more
effective in transferring entanglement to qubits than TWB states and
this opens novel perspectives on the use of non Gaussian states in
quantum information processing.\\
\indent The paper is organized as follows: in the next Section we
introduce the Hamiltonian model we are going to analyze for
entanglement transfer, as well as the different kind of two-mode CV
states that provide the source of entanglement. In Section
\ref{s:res} we consider resonant entanglement transfer, which is
assessed by evaluating the entanglement of formation for the reduced
density matrix of the qubits after a given interaction time. In
Sections \ref{s:det} and \ref{s:dift} we analyze in some details the
effects of detuning and of different interaction times for the two
qubits. Section \ref{s:out} closes the paper with some concluding
remarks.
\section{The Hamiltonian model}
We address the entanglement transfer from a bipartite CV field to a pair of
localized qubits assuming that each CV mode couples to one qubit via the
off-resonance Jaynes-Cummings interaction (as it happens by injecting the
two modes in two separate cavities). We allow for different interaction
times for the two subsystems and assume \cite{Zou 2006} that the initial
state of the two modes is described by a Schmidt decomposition in the
Fock number basis:
\begin{equation}
\label{CV}|x \rangle=\sum^{\infty}_{n=0}c_{n}(x)|nn\rangle
\end{equation}
where $|nn\rangle=|n\rangle\otimes|n\rangle$ and the complex
coefficients $c_n(x)=\langle n n|x\rangle$ satisfy the
normalization condition $\sum^{\infty}_{n=0}|c_{n}(x)|^2=1$. The
parameter $x$ is a complex variable that fully characterizes the
state of the field. Notice that a scheme for the generation of
any two-mode correlated photon number states of the form (\ref{CV})
has been recently proposed \cite{zou}.
The simplest example within the class (\ref{CV}) is
given by the Bell-like two-mode superposition (TSS):
\begin{equation}
\label{TSS} |x\rangle=c_{0}|00\rangle+c_{1}|11\rangle\:.
\end{equation}
Eq. (\ref{CV}) also describes relevant bipartite states, as for
example TWB and TMC states. In these cases we can rewrite the
coefficients as $c_n(x)=c_0(x)f_n(x)$, where:
\begin{align}
\label{TWB} {\rm TWB}:\qquad c_0(x)& =\sqrt{1-|x|^2} \quad f_n(x)=x^n\\
\label{PC}  {\rm TMC}:\qquad c_0(x)&= \frac{1}{\sqrt{I_0(2|x|)}}\quad
f_n(x)=\frac{x^n}{n!}\:,
\end{align}
where $I_0(y)$ denotes the $0$-th modified Bessel function of the
first kind. For TWB states the parameter $|x|$ is related to the
squeezing parameter, and ranges from 0 (no squeezing) to 1 (infinite
squeezing). For TMC states $|x|$ is related to the squared field
amplitude and can take any positive values. The bipartite states
described by (\ref{CV}) show perfect photon number correlations. The
joint photon number distribution has indeed the simple form
$P_{nk}(x) = \delta_{nk} |c_n(x)|^2$.  For the TSS states the joint
photon distribution is given by $P_{00}=|c_0|^2$ and
$P_{11}=1-P_{00}$, whereas for TWB and TMC it can be written as
$P_{nk}(x)=\delta_{nk} P_{00}(x)|f_n(x)|^2$. As we will see in the
following the photon distribution plays a fundamental role in
understanding the entanglement transfer process.
\par
\begin{figure}[h]
\begin{tabular}{cc}
a)\includegraphics[scale=0.26]{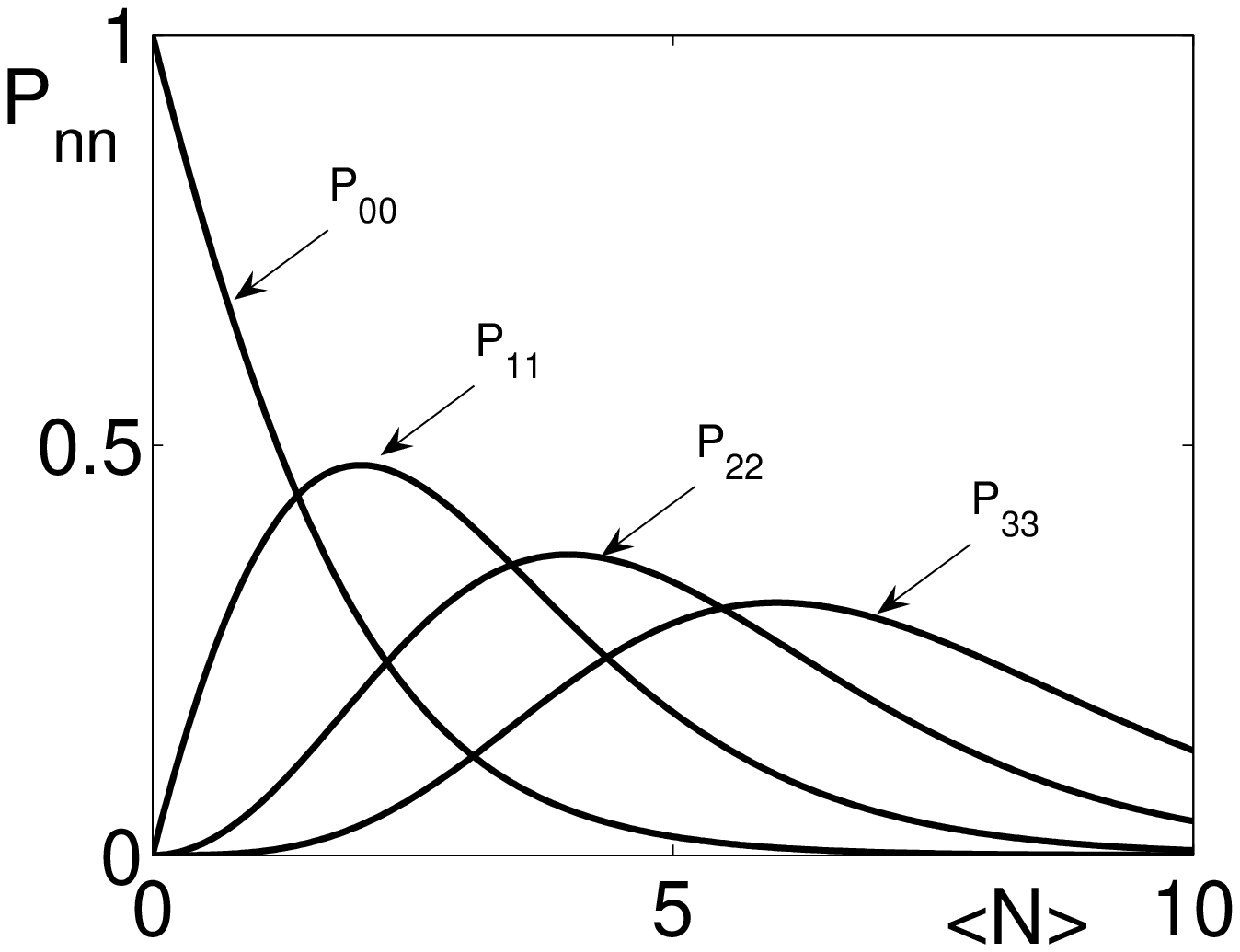} &
b)\includegraphics[scale=0.26]{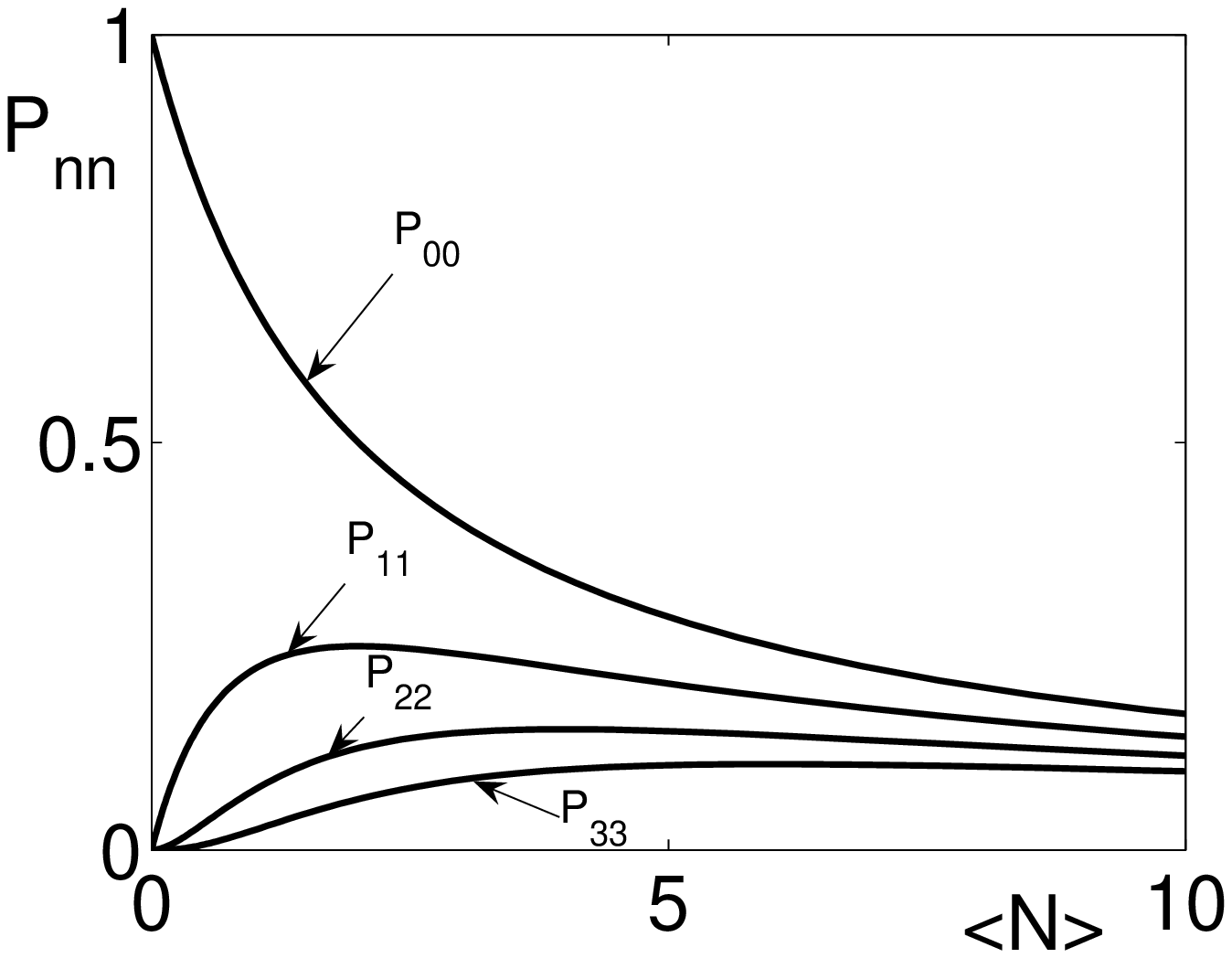}
\end{tabular}
\caption{The first four terms $P_{nn}$, $n=0,...,3$ of the joint
photon distribution of the state $|x\rangle$ as a function of
average photon number $\langle N\rangle$. (a): TMC, (b): TWB.}
\label{fig1}
\end{figure}

The average number of photon of the states $|x\rangle$, {\em i.e.}
$\langle N\rangle(x) = \langle x | a^\dag a + b^\dag b | x\rangle$,
$a$ and $b$ being the field mode operators, is related to the
dimensionless parameter $|x|$ by
\begin{align}
\label{MN}
{\rm TWB}: \qquad \langle N\rangle(x) & =2\frac{|x|^2}{1-|x|^2} \\
{\rm TMC}: \qquad \langle N\rangle(x)& =
\frac{2|x|I_1(2|x|)}{I_0(2|x|)}
\end{align}
where $I_1(y)$ denotes the $1$-th modified Bessel function of the
first kind. In Fig.\ref{fig1} we show the first four terms of the
photon distribution for TMC and TWB states respectively as functions of
the mean photon number.
\par
The states in Eq.~(\ref{CV}) are pure states, and therefore we can
evaluate their entanglement by the Von Neumann entropy $S_{vn}(x)$
of the reduced density matrix of each subsystem. For the TSS case we
simply have:
\begin{equation}
\label{BELLSTATE_SVN} S_{vn}=-P_{00}\log_2 P_{00}-(1-P_{00})\log_2
(1-P_{00})
\end{equation}
and of course the maximum value of 1 is obtained for
$P_{00}=P_{11}=\frac{1}{2}$. The corresponding state
$|\Phi_+\rangle=\frac{|00\rangle+|11\rangle}{\sqrt{2}}$ is a
Bell-like maximally entangled state.
For TWB states the Von Neumann Entropy can be written as:
\begin{eqnarray}
\label{SVNTWB} S_{vn}(x)=-\log_2
(1-|x|^2)-\frac{2|x|^2}{1-|x|^2}\log_2|x|
\end{eqnarray}
whereas  for TMC states we use the general expression:
\begin{eqnarray}
\label{SVN} S_{vn}(x)&=&-\log_2 P_{00}(x)-\,\nonumber\\
&-&P_{00}(x)\sum_{i=1}^{\infty} |f_i(x)|^2\log_2|f_i(x)|^2
\end{eqnarray}
It is clear that the Von Neumann entropy diverges in the limit
$|x|\rightarrow 1$ (for TWB) and in the limit $|x|\rightarrow
\infty$ (for TMC) because the probability $P_{00}(x)$ vanishes. The
VN entropy at fixed energy (average number of photons of the two
modes) is maximized by the TWB expression (\ref{SVNTWB}). For this
reason TWB states are also referred to as maximal entangled states
of bipartite CV systems.
\par
We consider the interaction of each radiation mode with a two level
atom flying through the cavity. If the interaction time is much
shorter than the lifetime of the cavity mode and the atomic decay
rates, we can neglect dissipation in system dynamics. On the other
hand, we consider the general case of atoms with different
interaction times and coupling constants, prepared in superposition
states, and off-resonance interaction between each atom and the
relative cavity mode. All these features can be quite important in
practical implementations such as in cavity QED systems with Rydberg
atoms and high-Q microwave cavities~\cite{Haroche}, as noticed
in~\cite{Paternostro2 2004}. In the interaction picture, the
interaction Hamiltonian $H_i$ is given by:
\begin{eqnarray}
\label{HI} H_{i}&=&-\hbar \Delta_A
a^{\dagger}a -\hbar \Delta_B b^{\dagger}b+\,\nonumber\\
&+&\hbar g_A [a^{\dagger}S^{12}_{A,-}+a S^{12}_{A,+}]+\hbar
g_B [b^{\dagger}S^{12}_{B,-}+bS^{12}_{B,+}]\,\nonumber\\
\end{eqnarray}
where $S^{12}_{A,\pm}$ and $S^{12}_{B,\pm}$ are the lowering and
raising atomic operators of the two atoms and $\Delta_A$, $\Delta_B$
denote the detunings between each mode frequency and the
corresponding atomic transition frequency. The initial state of the
whole system
$$|\psi(0)\rangle=|x\rangle\otimes|\psi(0)\rangle_{A}\otimes|\psi(0)\rangle_{B}$$
evolves by means of the unitary operator
$U(\tau)=\exp[-\frac{i}{\hbar}H_i\tau]$ that can be factorized as
the product of two off-resonance Jaynes-Cummings evolution operators
$U_A(\tau)$ and $U_B(\tau)$~\cite{ORSZAG} related to each atom-mode
subsystem. For the initial state of both atoms we considered a
general superposition of their excited ($|2\rangle$) and ground
($|1\rangle$) states:
\begin{eqnarray}
\label{ATOMstate}
|\psi(0)\rangle_{A}&=&A_2|2\rangle_{A}+A_1|1\rangle_{A}\,\nonumber\\
|\psi(0)\rangle_{B}&=&B_2|2\rangle_{B}+B_1|1\rangle_{B}
\end{eqnarray}
where $|A_2|^2+|A_1|^2=1$ and $|B_2|^2+|B_1|^2=1$. This includes the
most natural and widely investigated choice of both atoms in the
ground states, but will also allow us to investigate the effect of
different interaction times.\\
\indent Due to the linearity of evolution operator $U(\tau)$ and its
factorized form, the whole system state $|\psi(\tau)\rangle$ at a
time $\tau$ can be written as
\begin{eqnarray}
\label{EVOLVED STATE}
&&|\psi(\tau)\rangle=\,\nonumber\\
&&=c_0(x)\sum_{n=0}^{\infty}f_n(x)U_A(\tau)|\psi^n(0)\rangle_{A}\otimes
U_B(\tau)|\psi^n(0)\rangle_{B}\,\nonumber\\
\end{eqnarray}
where
$|\psi^n(0)\rangle_{A,B}=|\psi(0)\rangle_{A,B}\otimes|n\rangle_{A,B}$.
In each of two atom-field subspaces A and B we expand the
wave-function on the basis
$\{|2\rangle|k\rangle,|1\rangle|k+1\rangle\}^{\infty}_{k=0}\cup
\{|0\rangle|1\rangle\}$. The coefficients
$c_{A,1,k}(0)=_{A}\langle2|_{A}\langle k||\psi^n(0)\rangle_{A}$ and
$c_{B,1,k}(0)=_{B}\langle2|_{B}\langle k||\psi^n(0)\rangle_{B}$ of
the initial states are:
\begin{equation}
\label{INITCOEFF}\begin{array}{cc}
c_{A,1,0}(0)=A_1\delta_{n,0} & c_{B,1,0}(0)=B_1\delta_{n,0}  \\
c_{A,2,k}(0)=A_2\delta_{k,n} &c_{B,2,k}(0)=B_2\delta_{k,n} \\
c_{A,1,k+1}(0)=A_1\delta_{k+1,n} & c_{B,1,k+1}(0)=B_1\delta_{k+1,n}  \\
\end{array}
\end{equation}
The Jaynes-Cummings interaction couples only the coefficients of
each variety  $K$ whereas $c_{A,1,0}(0)$, $c_{B,1,0}(0)$ do not evolve.
Therefore, for each variety in the subspaces A and B the evolved
coefficients can be obtained by applying the off-resonance
Jaynes-Cummings $2\times2$ matrix $U_{jk}$ so that:
\begin{eqnarray}
\label{EVOLVCOEFF}
c_{2,k}(\tau)&=&U_{11}(k,\tau)c_{2,k}(0)+U_{12}(k,\tau)c_{1,k+1}(0)\,\nonumber\\
c_{1,k+1}(\tau)&=&U_{21}(k,\tau)c_{2,k}(0)+U_{22}(k,\tau)c_{1,k+1}(0)\,\nonumber
\end{eqnarray}
where
\begin{eqnarray}
\label{JCMATRIX}
U_{11}(k,\tau)&=&\cos(\frac{R_{k}\tau}{2})-
i\frac{\Delta}{R_{k}}\sin(\frac{R_{k}\tau}{2})\,\nonumber\\
U_{12}(k,\tau)&=&-
\frac{2ig\sqrt{k+1}}{R_{k}}\sin(\frac{R_{k}\tau}{2})=U_{21}(k,\tau),\nonumber\\
U_{22}(k,\tau)&=&\cos(\frac{R_{k}\tau}{2})+i\frac{\Delta}{R_{k}}\sin(\frac{R_{k}
\tau}{2})\,\nonumber\\
\end{eqnarray}
where the generalized Rabi frequencies are
$R_{k}=\sqrt{4g^2(k+1)+\Delta^2}$. To derive the evolved atomic
density operator $\rho_a^{1,2}$ we first consider the statistical
operator of the whole system
$\rho(\tau)=|\psi(\tau)\rangle\langle|\psi(\tau)|$ and then we trace
out the field variables. The explicit expressions of the density matrix
elements in the standard basis
$\{|2\rangle_A|2\rangle_B,|2\rangle_A|1\rangle_B,|1\rangle_A|2\rangle_B,|1
\rangle_A|1\rangle_B\}$ are reported in Appendix A.
\section{Entanglement transfer at resonance} \label{s:res}
\begin{figure}[h]
\begin{tabular}{c}
a)\includegraphics[scale=0.26]{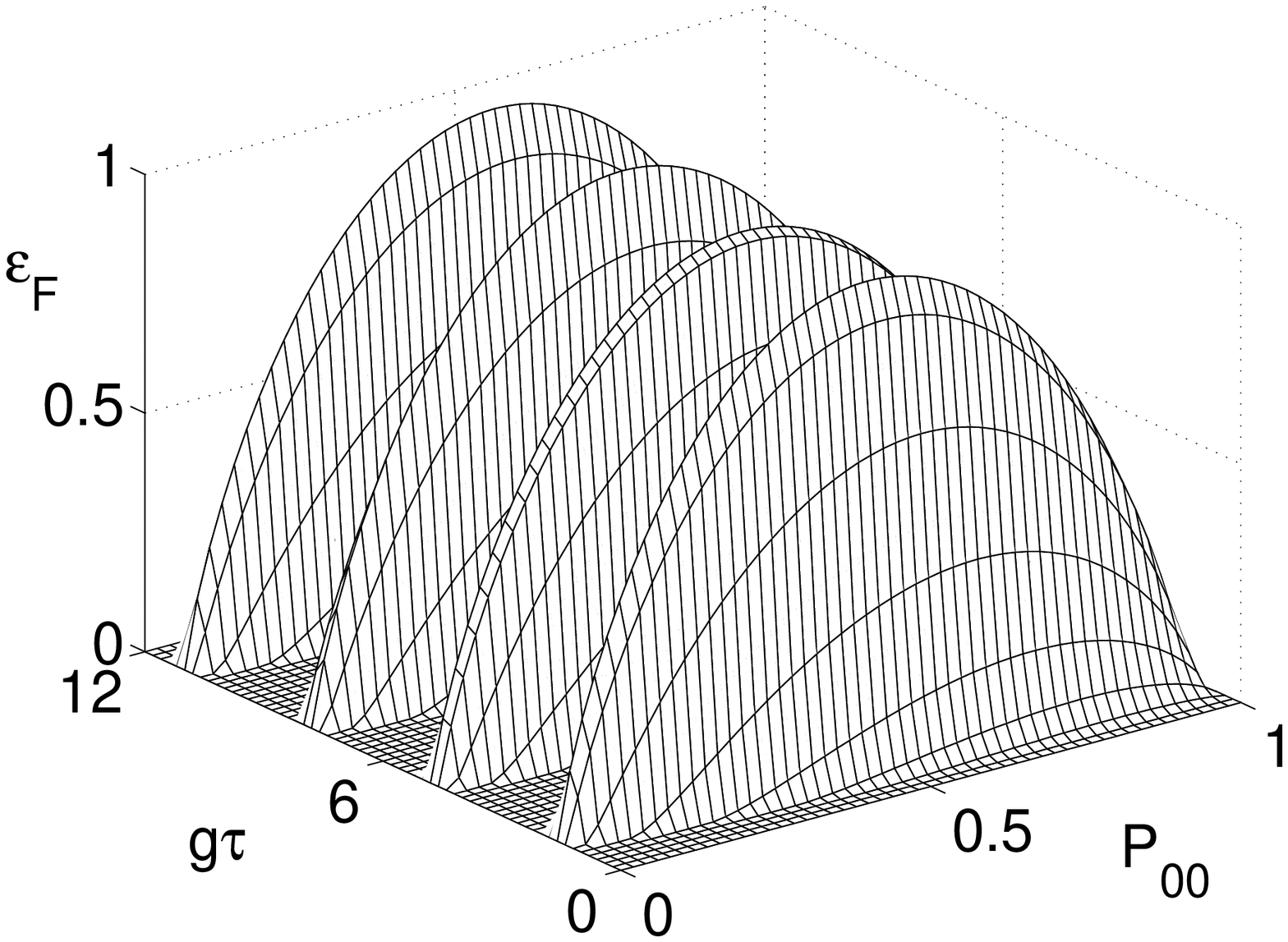}\\
b)\includegraphics[scale=0.26]{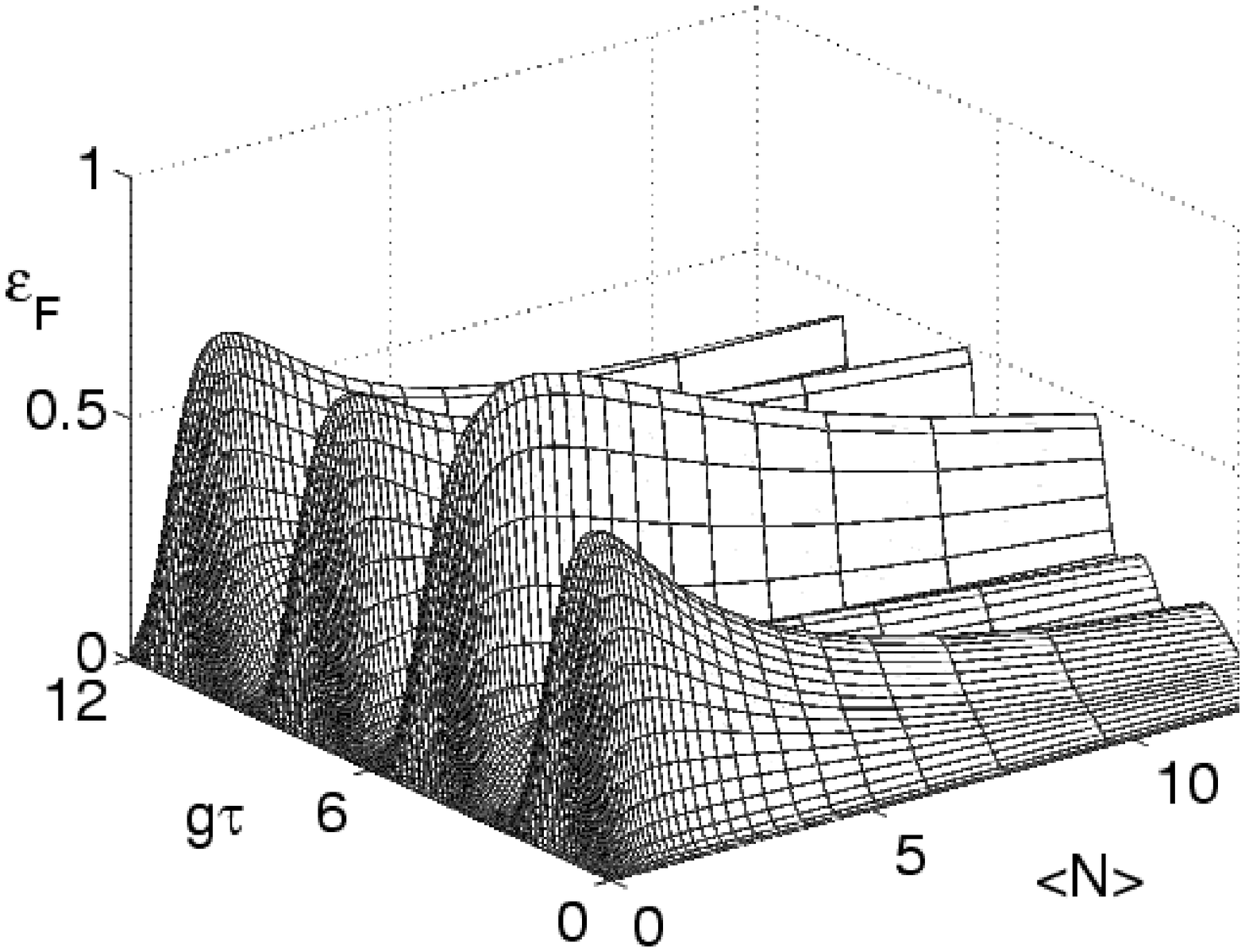}\\
c)\includegraphics[scale=0.26]{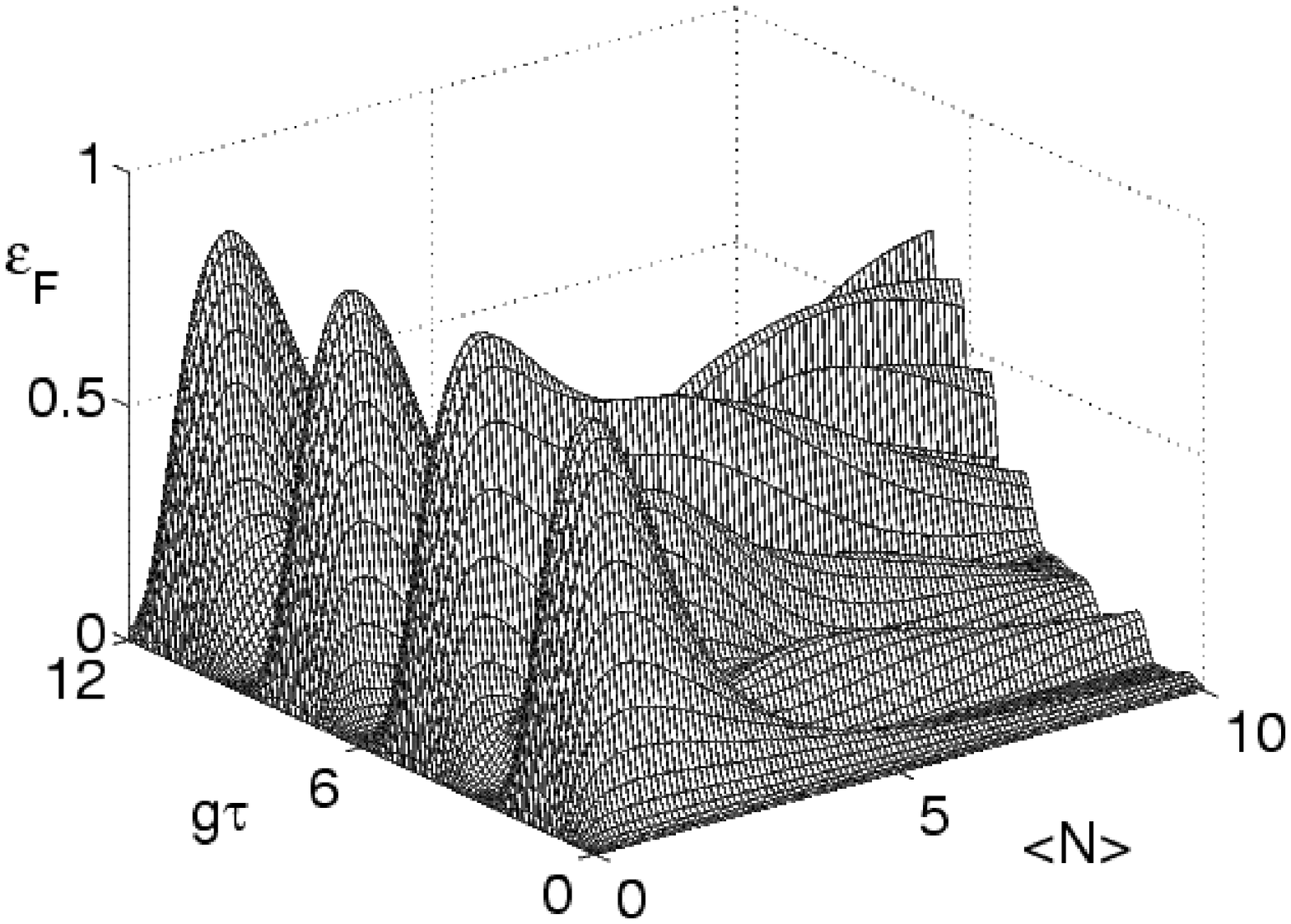}\\
\end{tabular}
\caption{Entanglement of formation $\epsilon_F$ of the qubit systems
as a function of the dimensionless time $g\tau$ and the CV state
parameter $P_{00}$ (a) or the average number of photons $\langle N\rangle$
(b,c) for the case of both atoms initially in the ground state.
a) TSS, b) TWB, c) TMC.} \label{fig2}
\end{figure}

As a first example we consider exact resonance for both atom-field
interactions, equal coupling constant $g$ and the same interaction
time $\tau$. For the initial atomic states we will discuss the
following three cases: both atoms in the ground state
($|1\rangle_A|1\rangle_B$), both atoms in the excited state
($|2\rangle_A|2\rangle_B$), and one atom in the excited state and
the other one in the ground state ($|1\rangle_A|2\rangle_B$). In all
these cases the atomic density matrix after the interaction
$\rho_a^{1,2}$ has the
following form:
\begin{equation}
\label{ROA12_case22} \rho_a^{1,2}=\left(\begin{array}{cccc}
\rho_{11} & 0 & 0 &\rho_{14}\\
0 &\rho_{22} & 0 & 0\\
0 & 0 & \rho_{33} & 0\\
\rho_{14}^*& 0 & 0 & \rho_{44}\\
\end{array}\right)
\end{equation}
The presence of the qubit entanglement can be revealed by the
Peres-Horodecki criterion~\cite{peres} based on the existence of
negative eigenvalues of the partial transpose of
Eq.~(\ref{ROA12_case22}). From the expressions of the eigenvalues
\begin{eqnarray}
\label{ROAeigen_case22}
&&\lambda_1^{PT}=\rho_{44}\hspace{1cm}\lambda_2^{PT}=\rho_{11}\hspace{1cm}\,
\nonumber\\
&&\lambda_{3,4}^{PT}=\frac{\rho_{22}+\rho_{33}\pm\sqrt{(\rho_{22}-
\rho_{33})^2+4|\rho_{14}|^2}}{2}
\end{eqnarray}
we see that only $\lambda_{4}^{PT}$ can assume negative values. In
the case of TSS the expression of $\lambda_{4}^{PT}$ can allow us to
derive in a simple way analytical results for the conditions of
maximum entanglement transfer as function of dimensionless
interaction time $g\tau$, as well as to better understand the
results in the case of TWB and TMC states. In order to quantify the
amount of the entanglement and, in turn, to assess the entanglement
transfer we choose to adopt the entanglement of formation
$\epsilon_F$~\cite{Bennet}. We rewrite the atomic density matrix in
the magic basis~\cite{Magic} $\rho_a^{MB}$ and we evaluate the
eigenvalues of the non-hermitian matrix
$R=\rho_a^{MB}(\rho_a^{MB})^*$:
\begin{equation}
\label{Reigen_case22}
\lambda_{1,2}^{R}=\rho_{22}\rho_{33}\hspace{1cm}\lambda_{3,4}^{R}=(\sqrt{\rho_{1
1}\rho_{44}}\pm|\rho_{14}|)^2
\end{equation}
In this way we calculate the concurrence~\cite{Wooters}
$C=max\{0,\Lambda_1-\Lambda_2-\Lambda_3-\Lambda_4\}$, where
$\Lambda_i=\sqrt{\lambda_i^R}$ are the square roots of the
eigenvalues $\lambda_{i}^{R}$ selected in the decreasing order, and
then evaluate the entanglement of formation:
\begin{eqnarray}
\label{EFdef}
\epsilon_{F}&=&-\frac{1-\sqrt{1-C^2}}{2}\log_2\frac{1-\sqrt{1-
C^2}}{2}\,\nonumber\\
&-&\frac{1+\sqrt{1-C^2}}{2}\log_2\frac{1+\sqrt{1-C^2}}{2}
\end{eqnarray}
\indent In the case of both qubits initially in the ground state
$|1\rangle_A|1\rangle_B$ the expression of $\lambda_{4}^{PT}$ simply
reduces to $\rho_{22}-|\rho_{14}|$, because $\rho_{22}=\rho_{33}$,
and it is possible to derive the following simple formula:
\begin{eqnarray}
\label{LPT4_case11}
&&\lambda_4^{PT}(P_{00},g\tau)=\sin^2(g\tau)\times\,\nonumber\\
&&\times[(1-P_{00})\cos^2(g\tau)-\sqrt{(1-P_{00})P_{00}}]
\end{eqnarray}
We note that only the vacuum Rabi frequency is involved, a fact that
greatly simplifies the analysis of atom-field interaction compared
to all the other atomic configurations. Let us first consider the
Bell-like state ($P_{00}=\frac{1}{2}$) and look for the $g\tau$
values maximizing the entanglement of the two atoms. The solution of
equation $\lambda_4^{PT}(\frac{1}{2},g\tau)=-\frac{1}{2}$ is given
by $g\tau=\frac{\pi}{2}(2k+1)$ with $k=0,1,2,...$. The above
condition is relevant also to explain the entanglement transfer for
TWB and TMC states, as discussed below. To evaluate the entanglement
transfer also for not maximally entangled TSS states in
Eq.~(\ref{TSS}), we calculate the entanglement of formation as a
function of both the dimensionless interaction time $g\tau$ and the
probability $P_{00}$. As it is apparent from Fig.~\ref{fig2}a there
are large and well defined regions where $\epsilon_F >0$. In
particular, the absolute maxima ($\epsilon_F=1$) occur exactly at
$P_{00}=0.5$ and for $g\tau$ values in agreement with the above
series. In addition, if we consider the sections at these $g\tau$
values, we obtain exact coincidence with the Von Neumann Entropy
function $S_{vn}(P_{00})$. Therefore, complete entanglement transfer
from the field to the atoms is possible not only for the Bell State,
though only for the Bell State we may obtain the transferral of 1
ebit.
In Fig.~\ref{fig2}b we consider the entanglement of formation vs
$g\tau$ and mean photon number $\langle N\rangle$ in the TWB case.
We first note that the regions of maximum entanglement correspond to
those of TSS states and the maxima occur at $g\tau$ values close to
$\frac{\pi}{2}(2k+1)$, as shown in~Table \ref{tab1}.
\begin{table}[h!]
\begin{tabular}{lllll}
\toprule
$g\tau_{max}$ & $\langle N\rangle_ {max}$ & $\epsilon_{F,max}$&$P_{00}$&$P_{11}$\\
\colrule
1.56 & 0.87 & 0.64 &0.69&0.21\\
4.61 & 1.82 & 0.81 &0.52&0.25\\
7.85 & 1.07 & 0.68 &0.65&0.23\\
11.03 &1.07 & 0.68 &0.65&0.23 \\
\botrule
\end{tabular}
\caption{Maxima of the qubit entanglement of formation $\epsilon_F$
for the resonant interaction with TWB and for both qubits initially in the
ground state (see Fig.~\ref{fig2}b).}
\label{tab1}
\end{table}
We can explain this by considering the TWB photon distribution (see
Fig.~\ref{fig1}b). We note that the terms $P_{00}$ and $P_{11}$ are
always greater or equal than the other terms $P_{nn}$ ($n>1$) and
that for $\langle N\rangle <2$ they dominate the photon distribution
($P_{00}+P_{11}\simeq1$). Therefore, the main contribution to
entanglement transfer is obtained from the above two terms as for
the TSS state. In order to explain the absolute maximum found in the
second peak at $\langle N\rangle=1.82$, we note that in this case
$P_{00}$ and $P_{11}$ are closer to the value $0.5$ of a Bell State.
In addition, for large $\langle N\rangle$ and $g\tau$ values, there
are small regions (not visible in the figure) where entanglement
transfer is possible. This is due to the terms $P_{nn}$ ($n>1$) in
the photon distribution. In Fig.~\ref{fig2}c we show the TMC case
and we note that for $\langle N\rangle <4$ there are four well
defined peaks where the entanglement is higher than in the TWB case.
Also in this case the $g\tau$ values of the maxima (see Table
\ref{tab2}) nearly correspond to those of TSS states.
\begin{table}[h!]
\begin{tabular}{lllll}
\toprule
$g\tau_{max}$ & $\langle N\rangle_ {max}$ & $\epsilon_{F,max}$&$P_{00}$&$P_{11}$\\
\colrule
1.56 & 0.89 & 0.84 & 0.61&0.34\\
4.66 & 1.09 & 0.90 & 0.54&0.39\\
7.85 & 0.99& 0.87 & 0.57&0.37\\
11.01 & 0.99& 0.88 & 0.57&0.37\\
\botrule
\end{tabular}
\caption{Maxima of the qubit entanglement of formation $\epsilon_F$
for the resonant interaction with TMC
and for both qubits initially in the ground state
(see Fig.~\ref{fig2}c).}\label{tab2}
\end{table}
As in the previous case this can be explained by the TMC photon
distribution (see Fig.~\ref{fig1}a), where for $\langle N\rangle<4$
the dominant components of the photon distribution are $P_{00}$ and
$P_{11}$. The absolute maximum is in the second peak at $\langle
N\rangle=1.09$ because $P_{00}$ and $P_{11}$ are even closer to the
Bell State than in the other peaks, and this also explains the
larger entanglement value $\epsilon_F$. For $\langle N\rangle>4$ and
large $g\tau$ there are regions with considerable entanglement
values, due to the fact that $P_{00}$ and $P_{11}$ are always
smaller than the other terms $P_{nn}$ ($n>1$) that dominate the
atom-field interaction.We note that the maxima of $\epsilon_F$ are
higher than in the TWB case.
\begin{figure}[h]
\begin{tabular}{cc}
a)\includegraphics[scale=0.26]{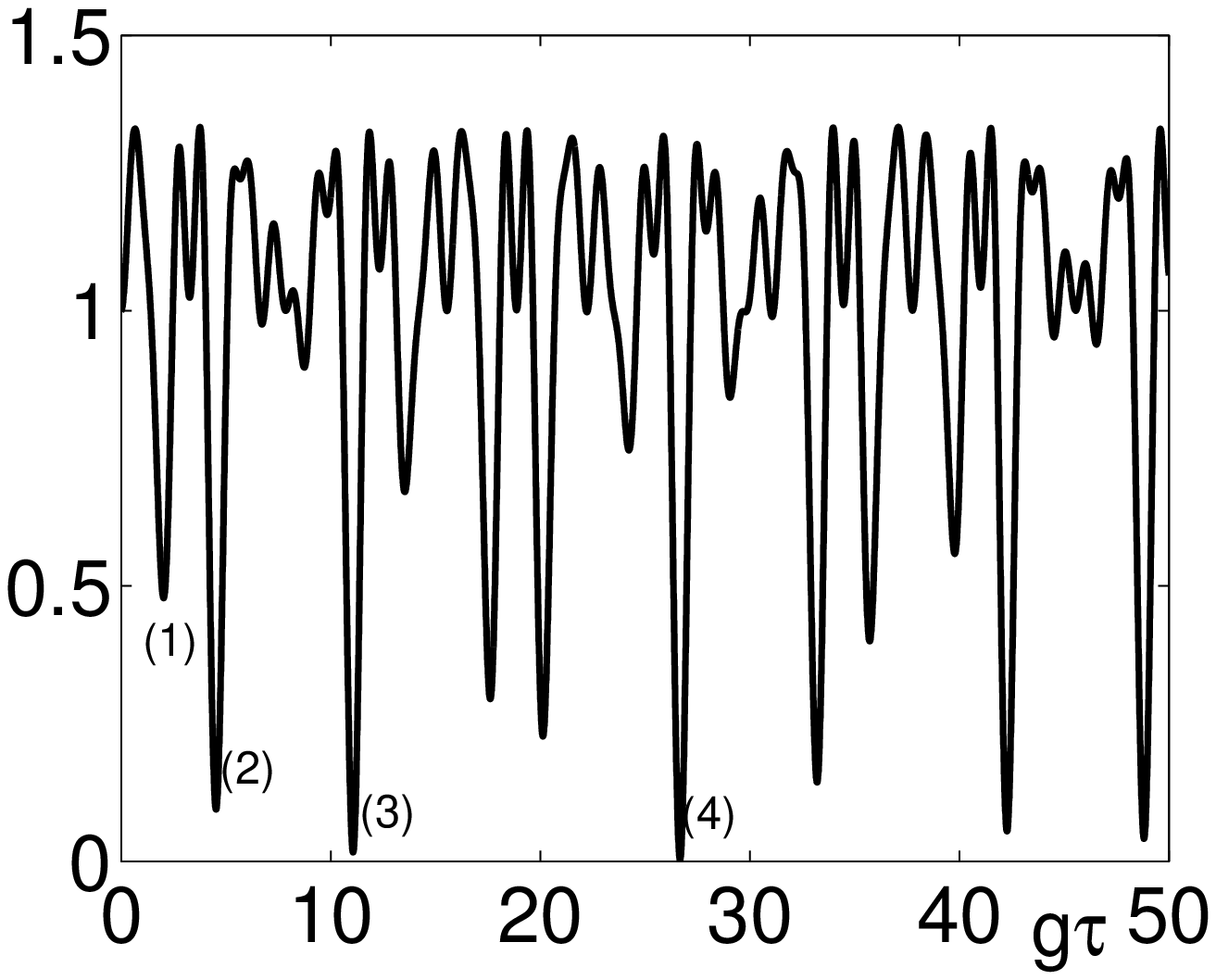} &
b)\includegraphics[scale=0.26]{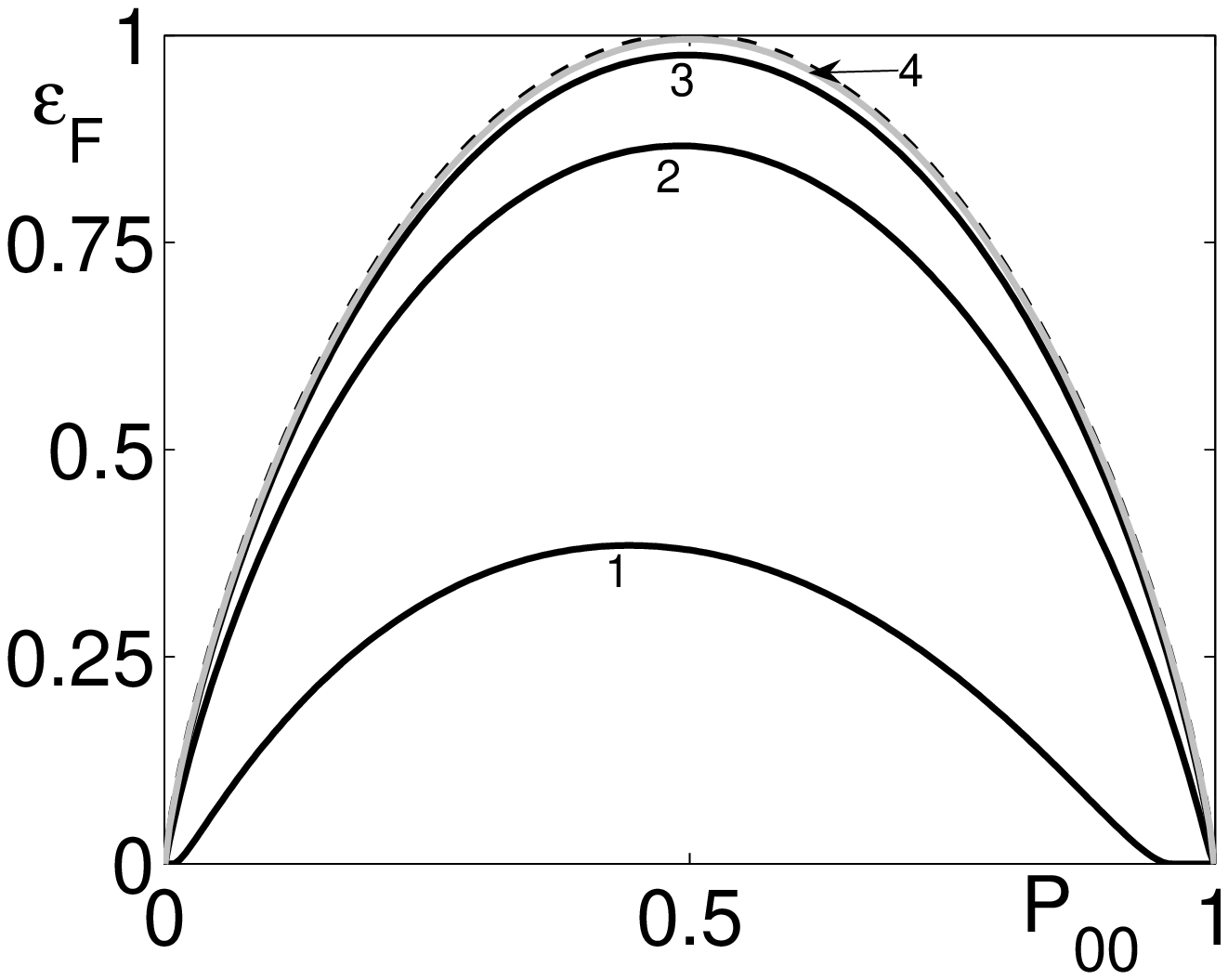}
\end{tabular}
\caption{(a): the function
$\lambda_4^{PT}(\frac{1}{2},g\tau)+\frac{1}{2}$ as in
Eq.~(\ref{LPT4_case22}) for the Bell state case with both atoms in
the excited state.  (b): entanglement of formation $\epsilon_F$ vs
$P_{00}$ for TSS states compared to Von Neumann entropy (dashed
line) for some values of $g\tau$, corresponding to the numbered
minima: (1) $2.03$, (2) $4.53$, (3) $11.07$, (4) $26.68$.}
\label{fig3}
\end{figure}
\par
A similar analysis can be done in the case of initially excited atoms
$|2\rangle_A|2\rangle_B$.
For the TSS states we can again write a simple equation for the
eigenvalue of the partial transpose:
\begin{eqnarray}
\label{LPT4_case22}
&&\lambda_4^{PT}(P_{00},g\tau)=(1-
P_{00})\sin^2(\sqrt{2}g\tau)\cos^2(\sqrt{2}g\tau)+\,\nonumber\\
&&+\sin^2(g\tau)[P_{00}\cos^2(g\tau)-\sqrt{(1-
P_{00})P_{00}}\cos^2(\sqrt{2}g\tau)]\,\nonumber\\
\end{eqnarray}
where, with respect to Eq.~(\ref{LPT4_case11}), an additional
frequency is present. For the Bell State we can look for $g\tau$
values maximizing the entanglement transfer. In this case the
problem can be solved numerically and we found, for example in the
range $g\tau=0-50$, that only for $g\tau=26.68\simeq\frac{17\pi}{2}$
we can solve the equation
$\lambda_4^{PT}(\frac{1}{2},g\tau)=-\frac{1}{2}$ with a good
approximation as shown in Fig.~\ref{fig3}a. In Fig.~\ref{fig3}b we
consider also non maximally entangled TSS states, showing the
entanglement of formation $\epsilon_F$ vs the probability $P_{00}$
for $g\tau$ values corresponding to numbered minima in
Fig.~\ref{fig3}a. We see that only for $g\tau=26.68$ a Bell State
can transfer 1 ebit of entanglement, but the entanglement transfer
is complete also for all the other $P_{00}$ values. A nearly
complete transfer can be obtained also for $g\tau=11.07$ but in the
other cases the entanglement transfer is only partial even for the
Bell State. We note that in~\cite{Zou 2006} it is shown that for
$g\tau=11.07$ one finds maximum entanglement transfer for both
atomic states $|1\rangle_A|1\rangle_B$, $|2\rangle_A|2\rangle_B$ but
starting with a different Bell-like field state
$|\Psi_{-}\rangle=\frac{|10\rangle-|01\rangle}{\sqrt{2}}$.
\par
In the TWB case we find large entanglement transfer for $g\tau$
values very close to the ones of minima (2-4) for the TSS states in
Fig.~\ref{fig3}a. Some $g\tau$ values corresponding to maxima of
$\epsilon_F$ in the case of both atoms in the ground state are
missing, and the best value of $\epsilon_F$ in the considered range
is at $g\tau=26.65$.
Also for the TMC states for small $\langle N\rangle$ we have large
entanglement transfer corresponding to the above $g\tau$ values,
but in addition for $\langle N\rangle>4$ and large $g\tau$ values
there are regions with considerable entanglement.\\
\indent Finally, in the case of one atom in the excited state and
the other one in the ground state ($|1\rangle_A|2\rangle_B$) it is
not possible to write for TSS states a simple equation as
Eq.~(\ref{LPT4_case22}) because in the atomic density matrix
Eq.~(\ref{ROA12_case22}) $\rho_{22}\neq \rho_{33}$ unlike in the
previous cases. However, there are again only two frequencies
involved as in the TSS case and we can do a similar analysis as
for both atoms in the excited state. Here we only mention that for
dimensionless interaction times corresponding to common maxima for
the different atomic states, the maxima of $\epsilon_F$ are rather
lower than in the cases ($|1\rangle_A|1\rangle_B$) and
($|2\rangle_A|2\rangle_B$), and more in general the transfer of
entanglement is sensibly reduced as a function of $\langle
N\rangle$.
\section{The detuning effect} \label{s:det}
\begin{figure}[h]
\begin{center}
a)\includegraphics[scale=0.25]{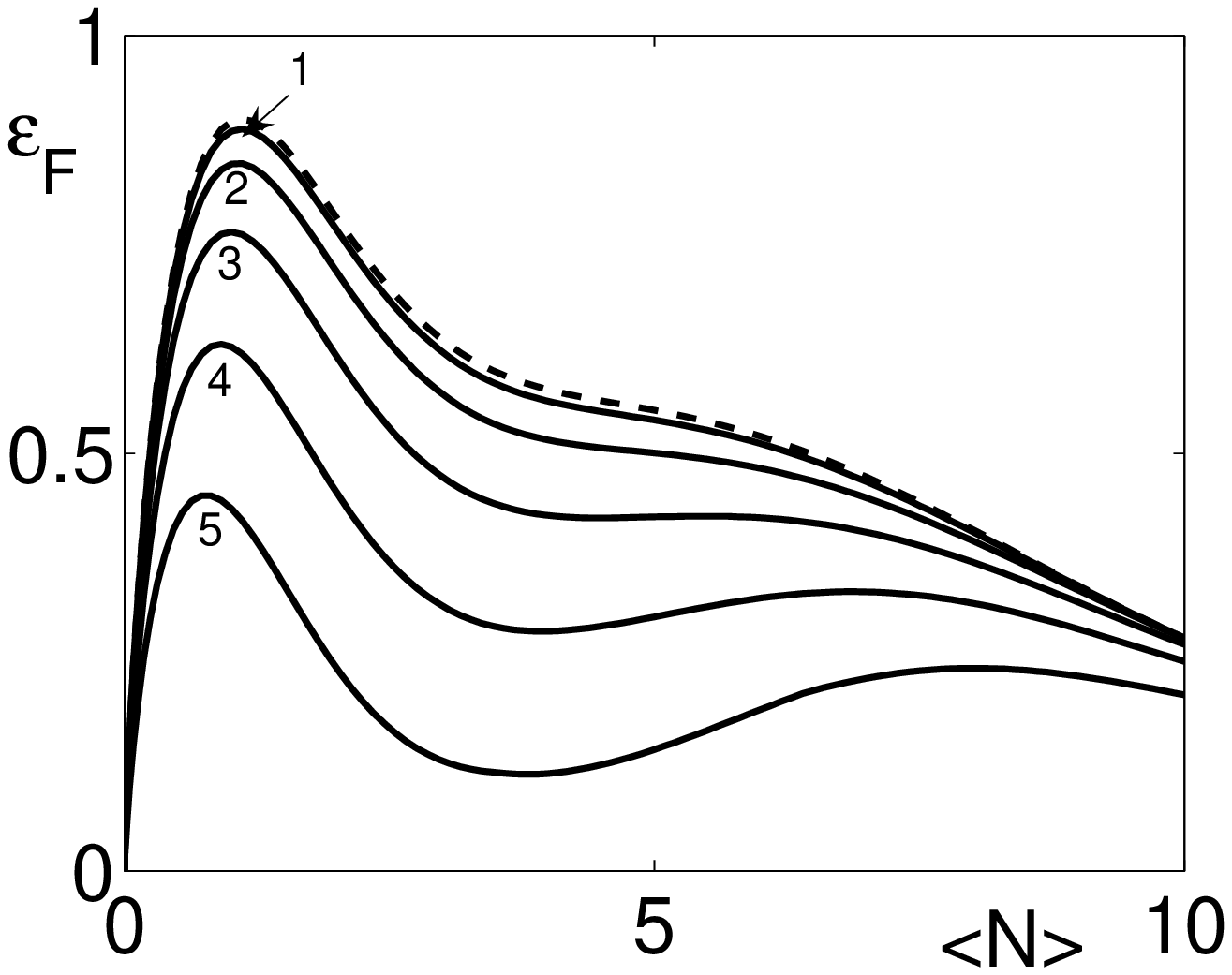}
b)\includegraphics[scale=0.25]{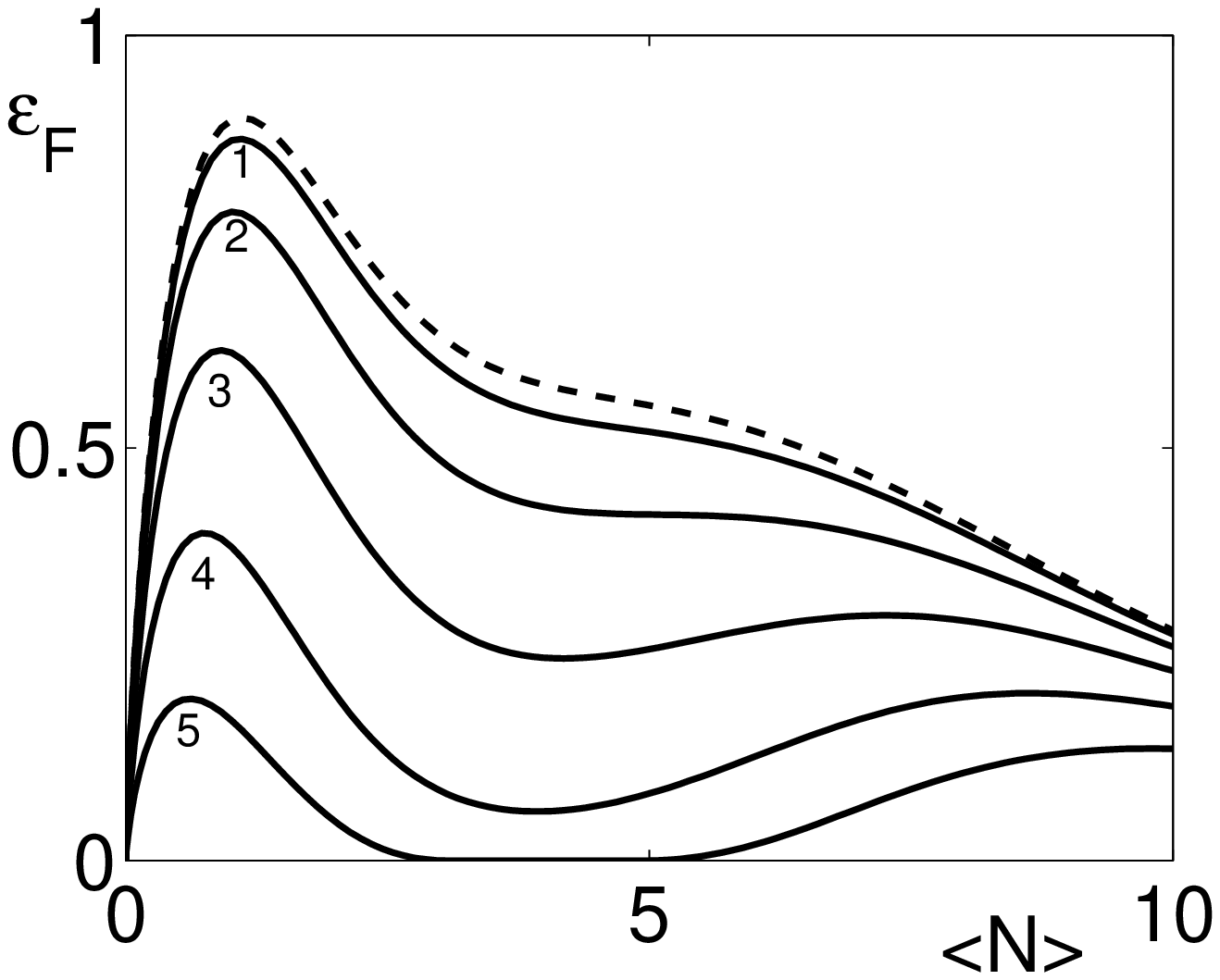}\\
\end{center}
\caption{The off resonance interaction effect in the TMC case, both
atoms in the ground state and $g\tau=4.66$. a) $\Delta_A\tau=0$, and
$\Delta_B\tau=0$ (dashed line), 1, 2, 3, 4, 5. b)
$\Delta_A\tau=\Delta_B\tau=0$ (dashed line), 1, 2, 3, 4, 5.}
\label{fig5}
\end{figure}

\begin{figure}[h]
\begin{center}
a)\includegraphics[scale=0.25]{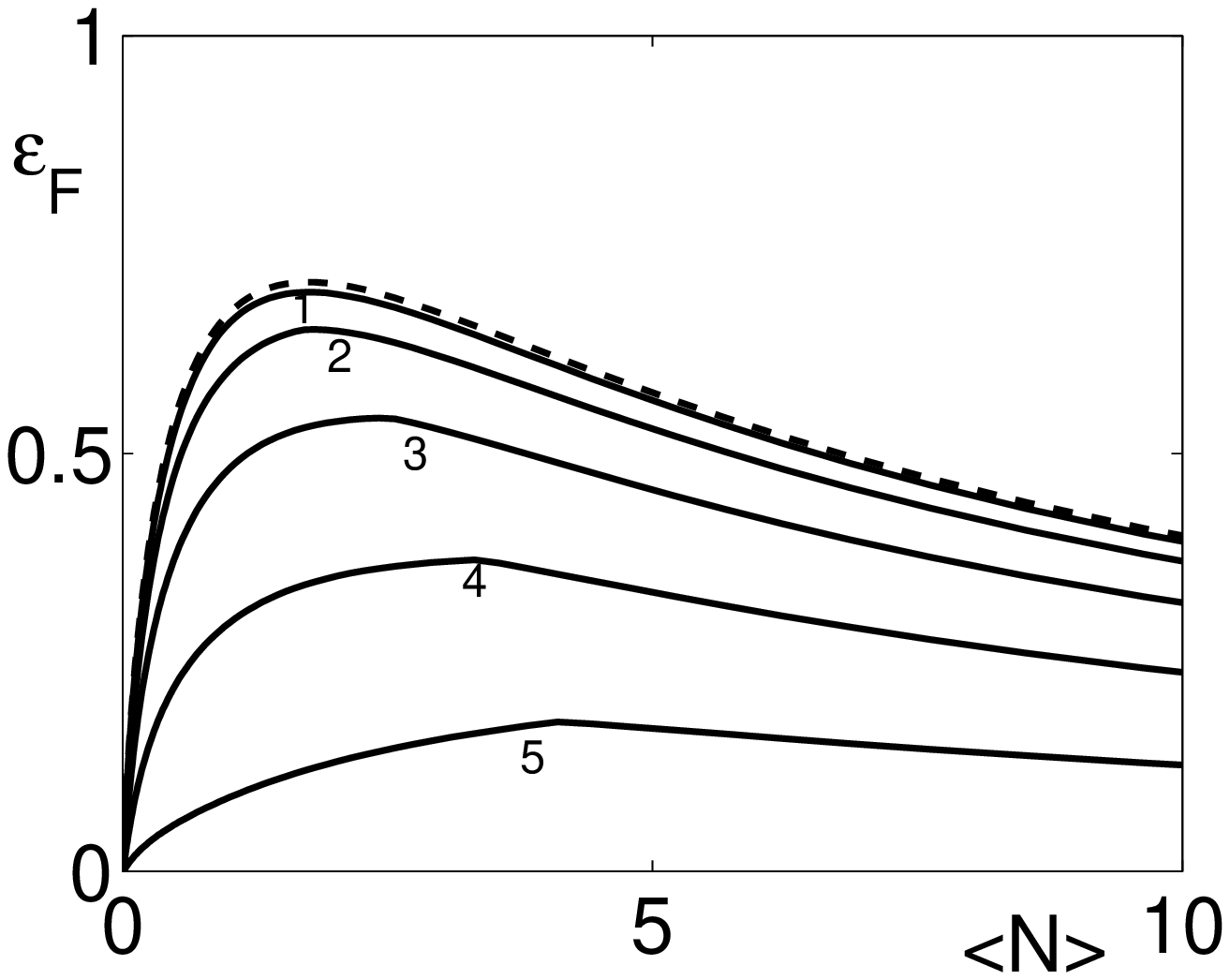}
b)\includegraphics[scale=0.25]{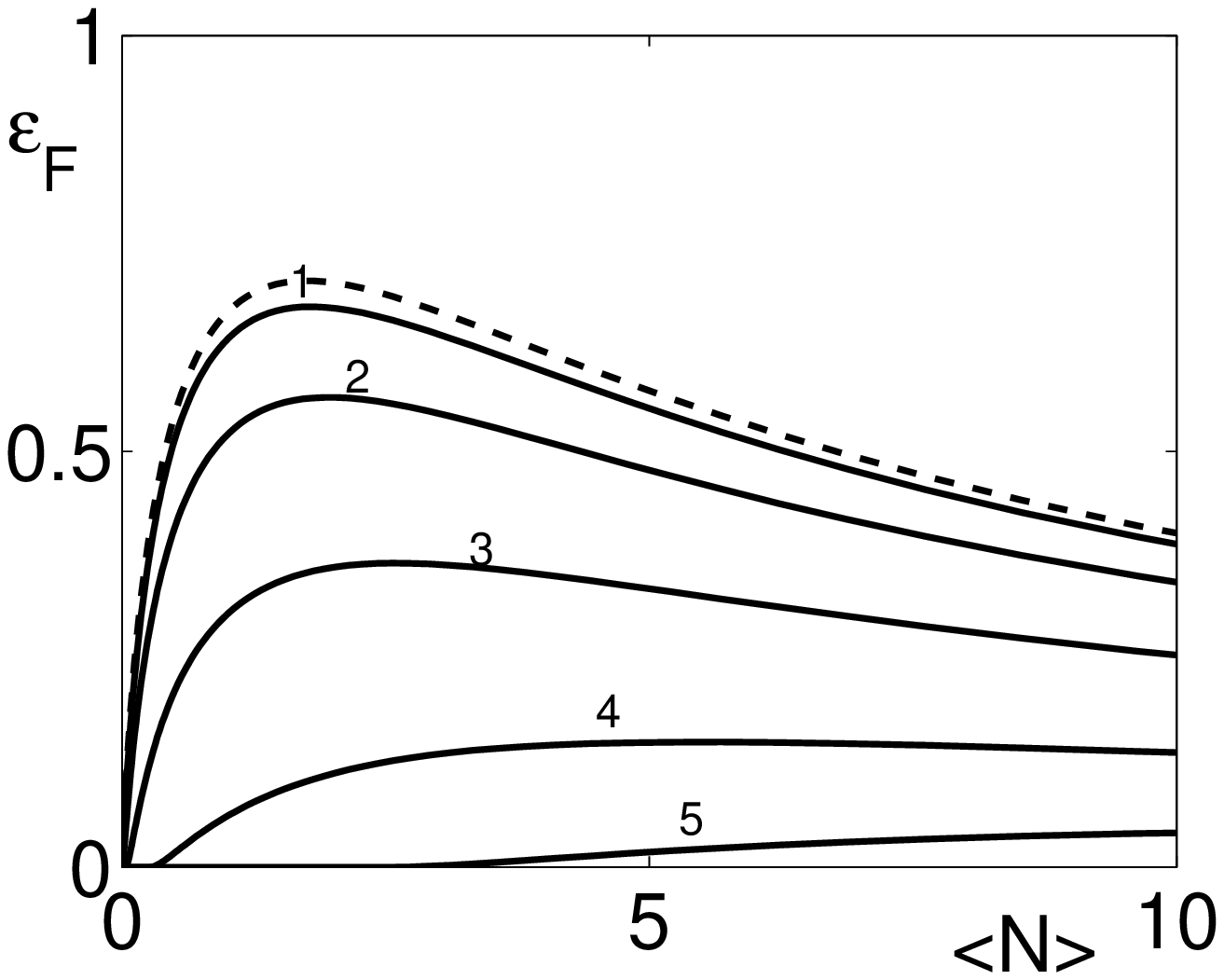}\\
\end{center}
\caption{The off resonance interaction effect in the TWB case, both
atoms in the ground state and $g\tau=4.61$. a) $\Delta_A\tau=0$, and
$\Delta_B\tau=0$ (dashed line), 1, 2, 3, 4, 5. b)
$\Delta_A\tau=\Delta_B\tau=0$ (dashed line), 1, 2, 3, 4, 5.}
\label{fig6}
\end{figure}

>From the practical point of view it is important to evaluate the
effects of the off-resonant interaction between the atoms and
their respective cavity fields, which can be actually prepared in
\textit{non degenerate} optical parametric processes. We assume
equal interaction times for both atoms and we consider first the
case of resonant interaction for the atom A and off resonant
interaction for atom B. As a first example we consider the TMC
case, both atoms in the ground state and the value $g\tau=4.66$,
corresponding to the maximum entanglement transfer. In
Fig.~\ref{fig5}a we see that up to detuning values on the order of
the inverse interaction time the entanglement is preserved by the
off-resonant interaction of atom B.
In Fig.~\ref{fig5}b we show the more general case of off-resonance for both
atoms,
taking equal detuning values for simplicity. The effect is greater than in
the previous case but it is negligible again up to $\Delta_B\tau=1$ .\\
Fig.~\ref{fig6} shows the analogous behaviour for the TWB states for
$g\tau=4.61$. We see that near the peak of entanglement the TMC
states seem more robust to off resonance interaction than the TWB
states.
\section{The effect of different interaction times}\label{s:dift}
\begin{figure}[h!]
\begin{center}
a)\includegraphics[scale=0.25]{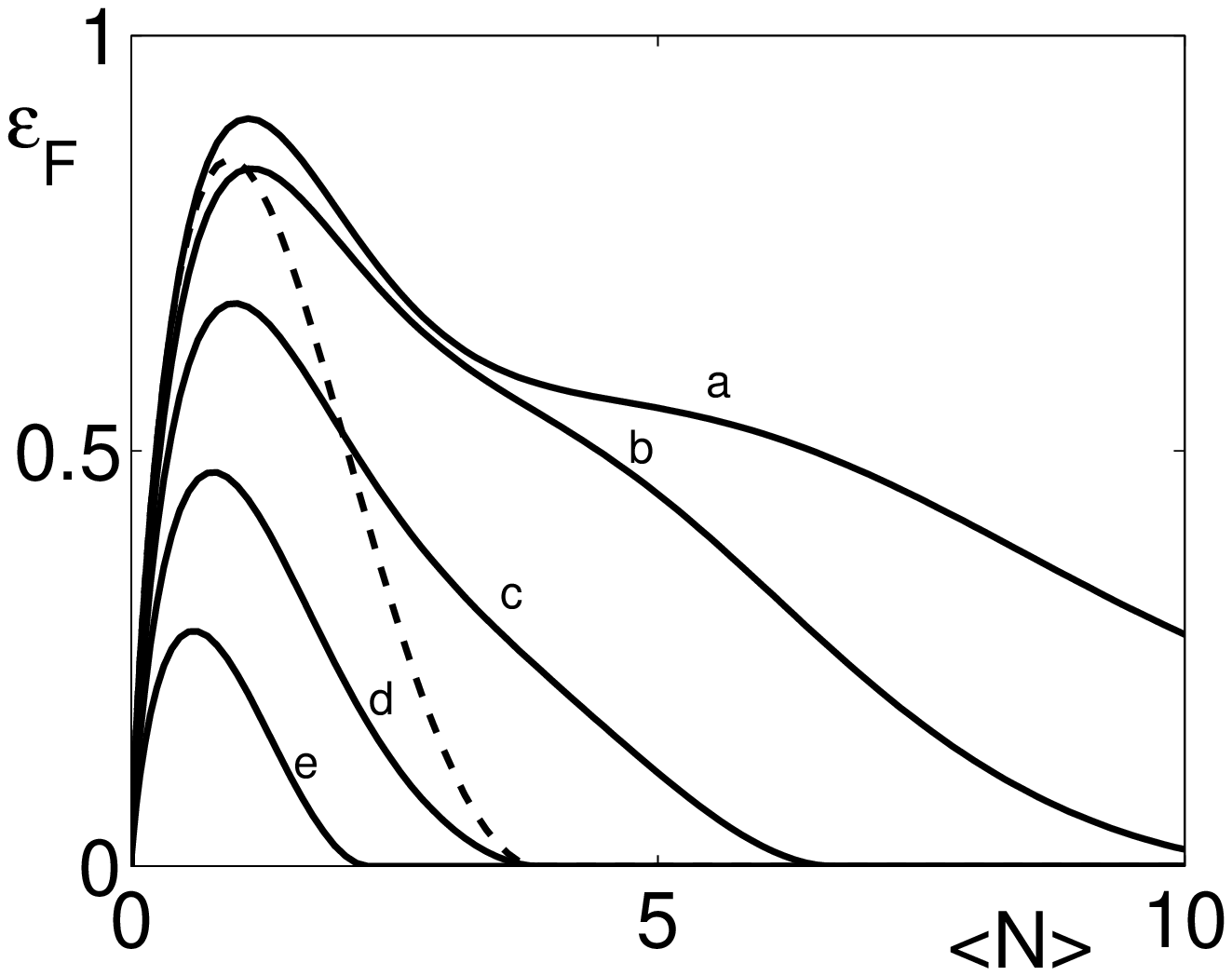}
b)\includegraphics[scale=0.25]{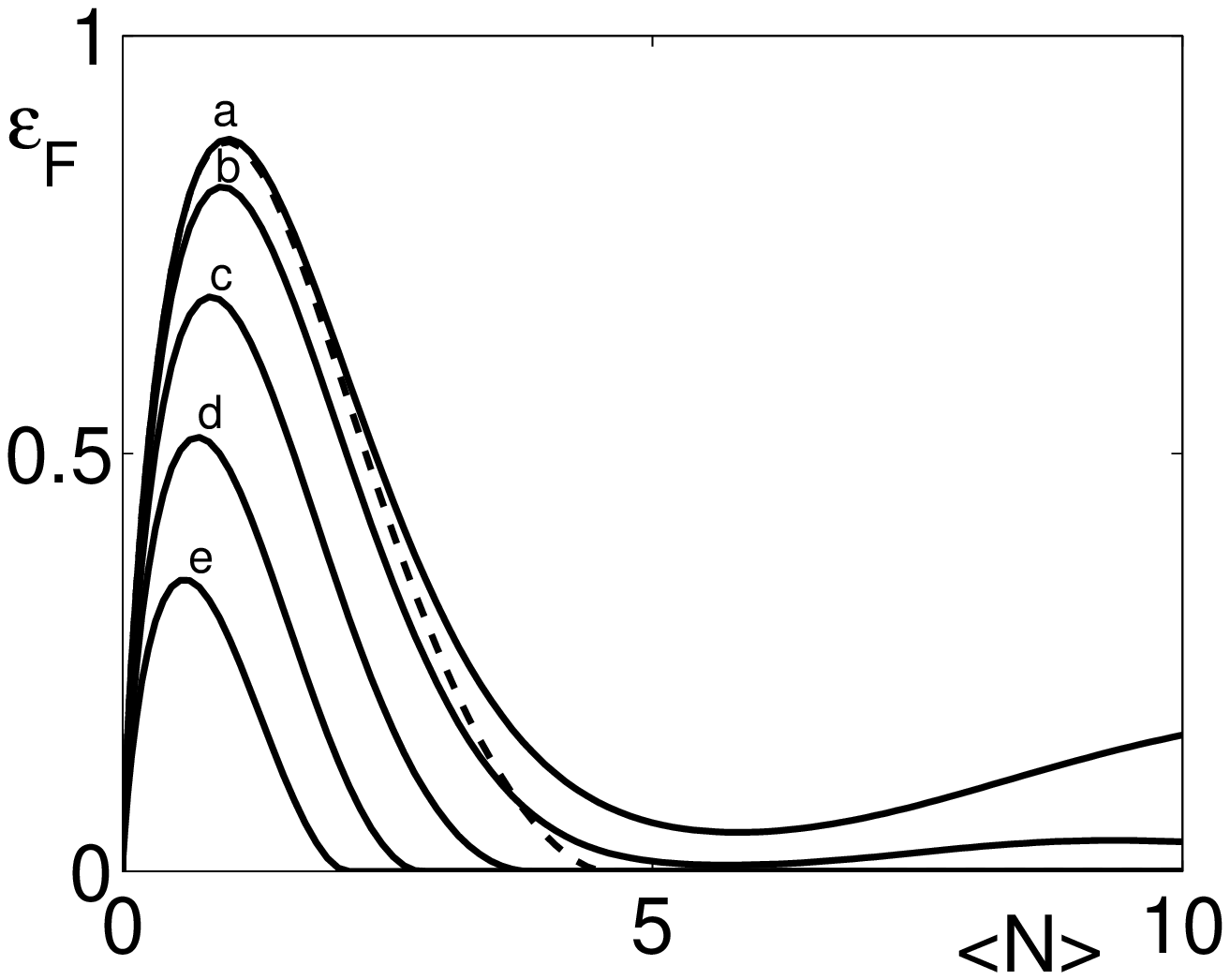} \\
c)\includegraphics[scale=0.25]{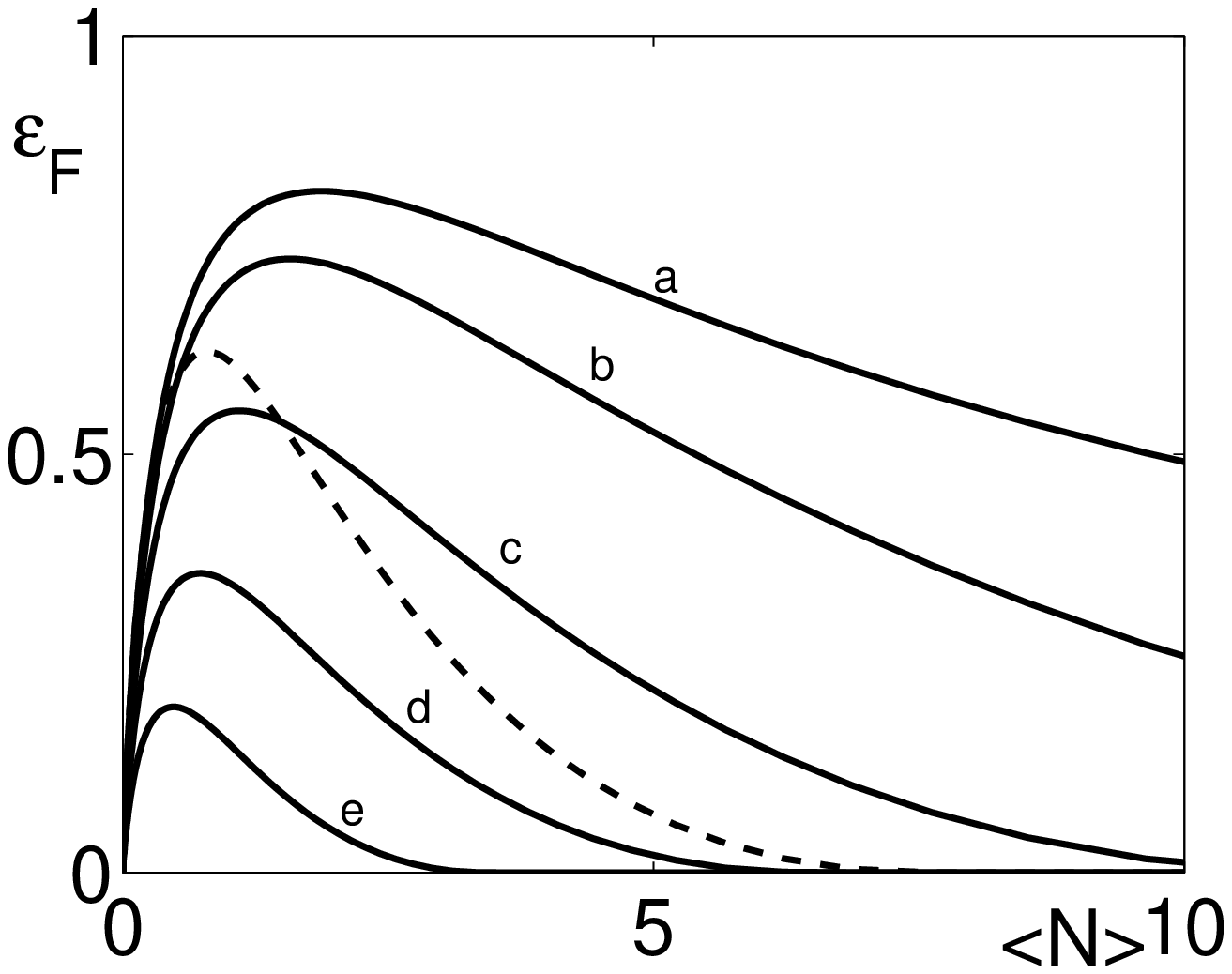}
d)\includegraphics[scale=0.25]{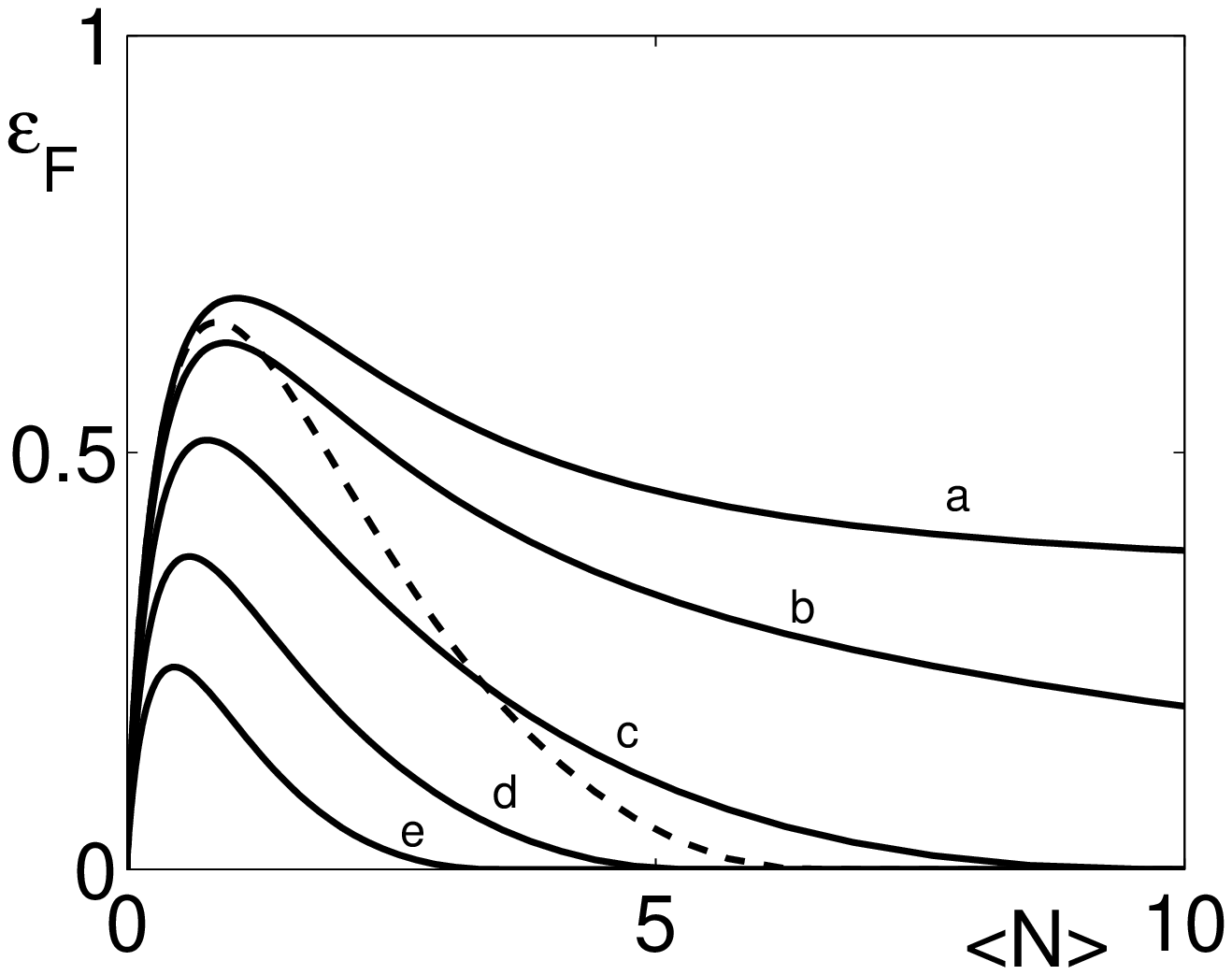}\vspace{-0.2cm}\\
\end{center}
\caption{The effect of different interaction times for
$\Delta_A=\Delta_B=0$, and both atoms injected simultaneously in the
ground state. In a) the TMC case for $g\tau_A=4.66$ and $g\tau_B$
values: a) 4.66, b) 4.4, c) 4.2, d) 4.0, e) 3.8, and 1.56 (dashed
line). In b) the TMC case for $g\tau_A=11.01$ and $g\tau_B$ values:
a) 11.01, b) 10.8, c) 10.6, d) 10.4, e) 10.2 and 7.85 (dashed line).
In c) the TWB case for $g\tau_A=4.61$ and $g\tau_B$ values: a) 4.61,
b) 4.4, c) 4.2, d) 4.0, e) 3.8, and 1.56 (dashed line). In d) the
TWB case for $g\tau_A=11.03$ and $g\tau_B$ values: a) 11.03, b)
10.8, c) 10.6, d) 10.4, e) 10.2 and 7.85 (dashed line).}
\label{fig7}
\end{figure}

\begin{figure}[h!]
\begin{center}
a)\includegraphics[scale=0.25]{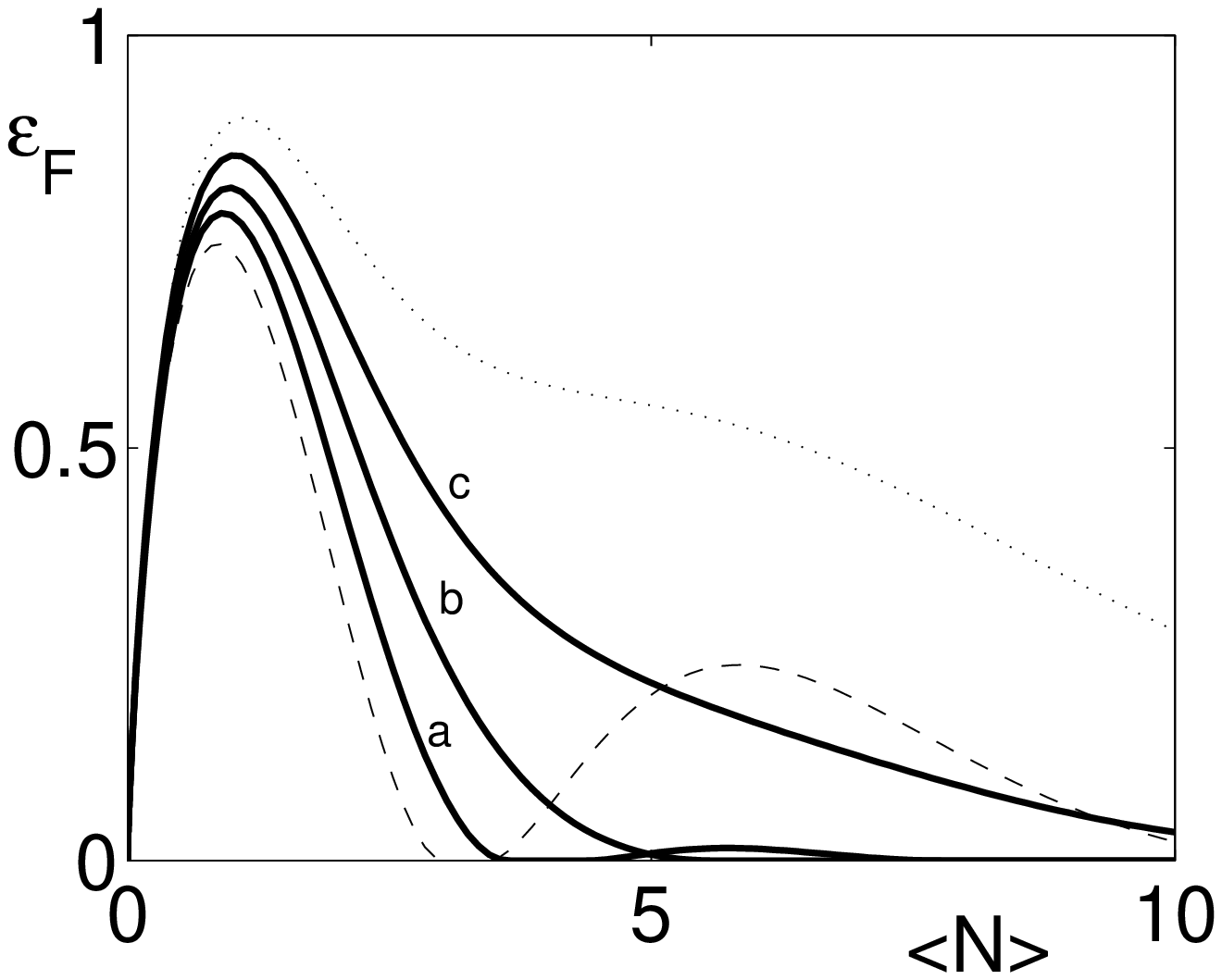}
b)\includegraphics[scale=0.25]{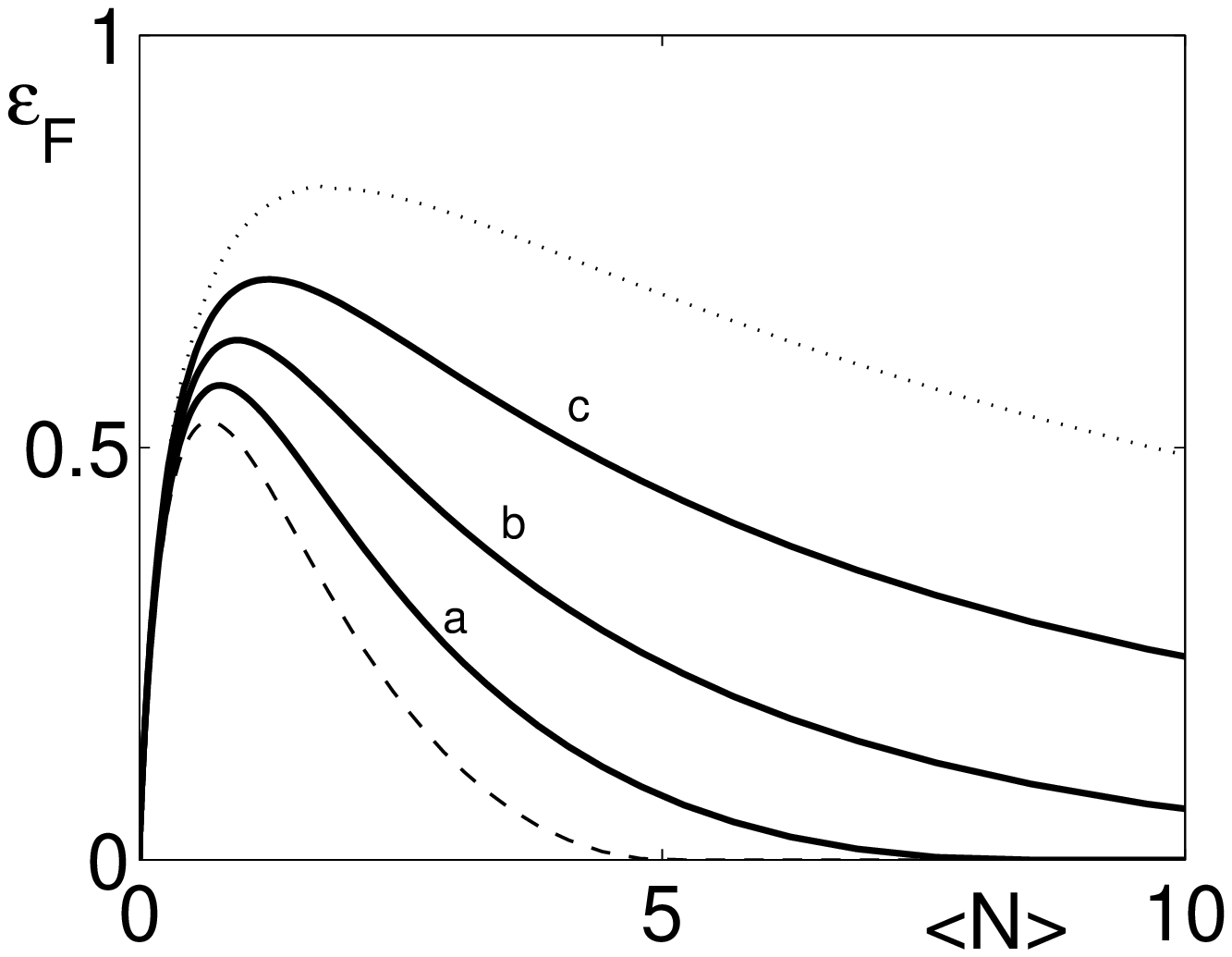}\\
\end{center}
\caption{The entanglement of formation $\epsilon_F$ vs $\langle
N\rangle$ for a time $g\tau$ of simultaneous presence of both atoms
after the delayed injection of atom A. Atom A is prepared in the
ground state and atom B in superposition states such that
$|B_1|^2=0$ (dashed line), 0.25 (a), 0.50 (b), 0.75 (c), 1 (dotted
line). a) the TMC case with $g\tau=4.66$. b) the TWB case with
$g\tau=4.61$.} \label{fig8}
\end{figure}

In the previous analysis we considered equal coupling constant and
interaction time for both atoms. However, experimentally we may
realize conditions such that the parameter $g\tau$ is different for
the two interactions due to the limitations in the control of both
atomic velocities and injection times or in the values of the
coupling constants~\cite{Paternostro2 2004}.\\
\indent We first consider the effect of different interaction times
at exact resonance and simultaneous injection of both atoms prepared
in the ground state. In Fig.~\ref{fig7}a,b we show the TMC case for
$g\tau_A=4.66$ and $g\tau_A=11.01$, corresponding to two maxima of
entanglement as discussed in the previous section, and we
investigate the effect of different dimensionless interaction times
for the atom B such that $g\tau_B\leq g\tau_A$. We see that
increasing the difference $g\tau_A-g\tau_B$ the entanglement
decreases.
However for $\langle N \rangle<4$, if $g\tau_B$ has a value close to
the one corresponding to a maximum, as for $g\tau_A=1.56$ in
Fig.~\ref{fig7}a and $g\tau_A=7.85$ in Fig.~\ref{fig7}b, i.e. if
$g\tau_A-g\tau_B\cong\pi$, the entanglement reaches again large
values. The effect is more important for $g\tau_A=11.01$ where the
entanglement transfer is the same as for equal interaction times. In
Fig.~\ref{fig7}c,d we show an analogous effect for the TWB case.
\indent We finally consider the possibility that atom B enters the
cavity just before atom A. We assume that when atom A enters its
cavity the two mode field can be still described by Eq.~(\ref{CV}).
Due to the interaction with its cavity field the atom B will be in a
superposition state $B_1|1\rangle_B+B_2|2\rangle_B$. In this case
the atomic density matrix $\rho_a^{12}$ has only two null elements,
$\rho_{23}=\rho_{32}^*$, hence we evaluate the eigenvalues of the
non-hermitian matrix numerically. We calculate the amount of
entanglement transferred to the atoms after the time $\tau$ of their
simultaneous presence into the respective cavities in the case of
exact resonance, equal coupling constant and velocity, assuming atom
A prepared in the ground state. In Fig.~\ref{fig8}a we show the TMC
case for $g\tau=4.66$ and different values of $|B_1|^2$ ranging from
0 (that is for $|1\rangle_A|2\rangle_B$) to 1 (that is
$|1\rangle_A|1\rangle_B$). We note that the behaviour of
$\epsilon_F$ gradually changes from one limit case to the other one
for $\langle N \rangle<4$, when the photon distribution approaches
that of a TSS state. For larger values of $\langle N \rangle$ we see
the occurrence of a second peak in the case
$|1\rangle_A|2\rangle_B$. In Fig.~\ref{fig8}b we show the TWB case
for $g\tau=4.61$ and we note that the gradual change described above
occurs for nearly all values of the mean photon number $\langle
N\rangle$.
\section{Conclusions}\label{s:out}
In this paper we have addressed the transfer of entanglement from a
bipartite state of a continuous-variable system to a pair of
localized qubits. We have assumed that each CV mode couples to one
qubit via the Jaynes-Cummings interaction and have taken into
account the degrading effects of detuning and of different
interaction times for the two subsystems. The transfer of
entanglement has been assessed by tracing out the field degrees of
freedom after the interaction, and then evaluating the entanglement
of formation of the reduced atomic density matrix.\\
\indent We found that CV states initially prepared in a two-state
superposition are the most efficient in transferring entanglement to
qubits with Bell-like states able to transfer a full ebit of
entanglement. We have then considered multiphoton preparation as TWB
and TMC states and found that there are large and well defined
regions of interaction parameters where the transfer of entanglement
is effective. At fixed energy (average number of photons) TMC states
are more effective in transferring entanglement than TWB states. We
have also found that the entanglement transfer is robust against the
fluctuations of interaction times and is not dramatically affected
by detuning. This kind of robustness is enhanced for the transfer of
entanglement from non Gaussian states as TMC states.\\
\indent  Overall, we conclude that the scheme analyzed in this paper
is a reliable and robust mechanism for the engineering of the
entanglement between two atomic qubits and that bipartite non
Gaussian states are promising resources in order to optimize this
protocol. Finally, we mention that our analysis may also be employed
to assess the entanglement transfer from radiation to
superconducting qubits.
\section*{Acknowledgments}
This work has been supported by MIUR through the project PRIN-2005024254-002.
MGAP thanks Vladylav Usenko for useful discussions about pair-coherent (TMC)
states.
\appendix
\begin{widetext}
\section{ATOMIC DENSITY MATRIX ELEMENTS}
The elements of the $4\times4$ atomic density matrix $\rho_a^{12}$
in the standard basis
$\{|2\rangle_A|2\rangle_B,|2\rangle_A|1\rangle_B,|1\rangle_A|2\rangle_B,|1
\rangle_A|1\rangle_B\}$
are listed below.
\begin{eqnarray}
\label{RHOA11APP}&&\rho_{11}(x)=|c_0(x)|^2\{|A_2|^2|B_2|^2\sum_{j=0}^{\infty}|f_j(x)|^2|U_{A11}(j,\tau)|^2|U_{B11}(j,\tau)|^2+\,\nonumber\\
&&+A_1^*A_2B_1^*B_2\sum_{j=0}^{\infty}f_j(x)f_{j+1}^*(x)U_{A11}(j,\tau)U_{B11}(j,\tau)U_{A12}^*(j,\tau)U_{B12}^*(j,\tau)+\,\nonumber\\
&&+|A_2|^2|B_1|^2\sum_{j=0}^{\infty}|f_{j+1}(x)|^2|U_{A11}(j+1,\tau)|^2|U_{B12}(
j,\tau)|^2+|A_1|^2|B_2|^2\sum_{j=0}^{\infty}|f_{j+1}(x)|^2|U_{A12}(j,\tau)|^2|U_
{B11}(j+1,\tau)|^2+\,\nonumber\\
&&+|A_1|^2|B_1|^2\sum_{j=0}^{\infty}|f_{j+1}(x)|^2|U_{A12}(j,\tau)|^2|U_{B12}(j,
\tau)|^2+\,\nonumber\\
&&+A_1A_2^*B_1B_2^*\sum_{j=0}^{\infty}
f_{j+1}(x)f_{j}^*(x)U_{A12}(j,\tau)U_{B12}(j,\tau)U_{A11}^*(j,\tau)U_{B11}^*(j,\
tau)\}
\end{eqnarray}
\begin{eqnarray}
\label{RHOA22APP_A} &&\rho_{22}(x)=|c_0(x)|^2
\{|A_2|^2|B_2|^2\sum_{j=1}^{\infty}|f_{j-1}(x)|^2|U_{A11}(j-
1,\tau)|^2|U_{B21}(j-1,\tau)|^2+\,\nonumber\\
&&+A_1^*A_2B_1^*B_2\sum_{j=1}^{\infty} f_{j-1}(x) f_{j}^*(x)U_{A11}(j-
1,\tau)U_{B21}(j-1,\tau)U_{A12}^*(j-1,\tau)U_{B22}^*(j-1,\tau)+\,\nonumber\\
&&+|A_2|^2|B_1|^2[\sum_{j=1}^{\infty}|f_{j}(x)|^2|U_{A11}(j,\tau)|^2|U_{B22}(j-
1,\tau)|^2+|U_{A11}(0,\tau)|^2]+\,\nonumber\\
&&+|A_1|^2|B_2|^2\sum_{j=1}^{\infty}|f_{j}(x)|^2|U_{A12}(j-
1,\tau)|^2|U_{B21}(j,\tau)|^2+|A_1|^2|B_1|^2\sum_{j=1}^{\infty}|f_{j}(x)|^2|U_{A
12}(j-1,\tau)|^2|U_{B22}(j-1,\tau)|^2+\,\nonumber\\
&&+A_1A_2^*B_1B_2^*\sum_{j=1}^{\infty} f_{j}(x)f_{j-1}^*(x)U_{A12}(j-
1,\tau)U_{B22}(j-1,\tau)U_{A11}^*(j-1,\tau)U_{B21}^*(j-1,\tau)\}
\end{eqnarray}
\begin{eqnarray}
\label{RHOA33APP_A} &&\rho_{33}(x)=|c_0(x)|^2
\{|A_2|^2|B_2|^2\sum_{j=0}^{\infty}|f_{j}(x)|^2|U_{A21}(j,\tau)|^2|U_{B11}(j,
\tau)|^2+\,\nonumber\\
&&+A_1^*A_2B_1^*B_2\sum_{j=0}^{\infty} f_{j}(x)
f_{j+1}^*(x)U_{A21}(j,\tau)U_{B11}(j,\tau)U_{A22}^*(j,\tau)U_{B12}^*(j,\tau)+\,
\nonumber\\
&&+|A_1|^2|B_2|^2[\sum_{j=1}^{\infty}|f_{j}(x)|^2|U_{A22}(j-
1,\tau)|^2|U_{B11}(j,\tau)|^2+|U_{B11}(0,\tau)|^2]+\,\nonumber\\
&&+|A_2|^2|B_1|^2\sum_{j=0}^{\infty}|f_{j+1}(x)|^2|U_{A21}(j+1,\tau)|^2|U_{B12}(
j,\tau)|^2+|A_1|^2|B_1|^2\sum_{j=0}^{\infty}|f_{j+1}(x)|^2|U_{A22}(j,\tau)|^2|
U_{B12}(j,\tau)|^2+\,\nonumber\\
&&+A_1A_2^*B_1B_2^*\sum_{j=0}^{\infty}
f_{j+1}(x)f_{j}^*(x)U_{A22}(j,\tau)U_{B12}(j,\tau)U_{A21}^*(j,\tau)U_{B11}^*(j,
\tau)\}
\end{eqnarray}
\begin{eqnarray}
\label{RHOA44APP_A} &&\rho_{44}(x)=|c_0(x)|^2
\{|A_2|^2|B_2|^2\sum_{j=1}^{\infty}|f_{j-1}(x)|^2|U_{A21}(j-
1,\tau)|^2|U_{B21}(j-1,\tau)|^2+\,\nonumber\\
&&+A_1^*A_2B_1^*B_2\sum_{j=1}^{\infty} f_{j-1}(x) f_{j}^*(x)U_{A21}(j-
1,\tau)U_{B21}(j-1,\tau)U_{A22}^*(j-1,\tau)U_{B22}^*(j-1,\tau)+\,\nonumber\\
&&+|A_2|^2|B_1|^2[\sum_{j=1}^{\infty}|f_{j}(x)|^2|U_{A21}(j,\tau)|^2|U_{B22}(j-
1,\tau)|^2+|U_{A21}(0,\tau)|^2]+\,\nonumber\\
&&+|A_1|^2|B_2|^2[\sum_{j=1}^{\infty}|f_{j}(x)|^2|U_{A22}(j-
1,\tau)|^2|U_{B21}(j,\tau)|^2+|U_{B21}(0,\tau)|^2]+\,\nonumber\\\
&&+|A_1|^2|B_1|^2[\sum_{j=1}^{\infty}|f_{j}(x)|^2|U_{A22}(j-
1,\tau)|^2+|U_{B22}(j-1,\tau)|^2+1]+\,\nonumber\\
&&+A_1A_2^*B_1B_2^*\sum_{j=1}^{\infty} f_{j}(x) f_{j-1}^*(x)U_{A22}(j-1,\tau)
U_{B22}(j-1,\tau)U_{A21}^*(j-1,\tau)U_{B21}^*(j-1,\tau)\}
\end{eqnarray}
\begin{eqnarray}
\label{RHOA12APP_A}&&\rho_{12}(x)=|c_0(x)|^2
\{|A_2|^2B_1^*B_2[\sum_{j=1}^{\infty}|f_{j}(x)|^2|U_{A11}(j,\tau)|^2U_{B11}(j,
\tau)U_{B22}^*(j-1,\tau)+|U_{A11}(0,\tau)|^2U_{B11}(0,\tau)]+\,\nonumber\\
&&+|A_1|^2B_2B_1^*\sum_{j=1}^{\infty}|f_{j}(x)|^2U_{A12}(j-
1,\tau)|^2|U_{B11}(j,\tau)U_{B22}^*(j-1,\tau)+\,\nonumber\\
&&+A_1A_2^*|B_2|^2\sum_{j=1}^{\infty}f_{j}(x)f_{j-1}^*(x) U_{A12}(j-
1,\tau)U_{B11}(j,\tau)U_{A11}^*(j-1,\tau)U_{B21}^*(j-1,\tau)+\,\nonumber\\
&&+A_1A_2^*|B_1|^2[\sum_{j=1}^{\infty}f_{j+1}(x)f_{j}^*(x)
U_{A12}(j,\tau)U_{B12}(j,\tau)U_{A11}^*(j,\tau)U_{B22}^*(j-
1,\tau)+f_{1}(x)U_{A12}(0,\tau)U_{B12}(0,\tau)U_{A11}^*(0,\tau)]\}\,\nonumber\\
\end{eqnarray}
\begin{eqnarray}
\label{RHOA13APP_A}&&\rho_{13}(x)=|c_0(x)|^2
\{|A_2|^2B_1B_2^*\sum_{j=0}^{\infty}f_{j+1}(x)f_{j}^*(x)U_{A11}(j+1,\tau)U_{B12}
(j,\tau)U_{A21}^*(j,\tau)U_{B11}^*(j,\tau)+\,\nonumber\\
&&+|A_1|^2B_1B_2^*[\sum_{j=1}^{\infty}f_{j+1}(x)f_{j}^*(x)U_{A12}(j,\tau)U_{B12}
(j,\tau)U_{A22}^*(j-
1,\tau)U_{B11}^*(j,\tau)+f_1(x)U_{A12}^(0,\tau)U_{B12}(0,\tau)U_{B11}^*(0,\tau)
]+\,\nonumber\\
&&+A_1^*A_2|B_2|^2[\sum_{j=1}^{\infty}|f_{j}(x)|^2
U_{A11}(j,\tau)|U_{B11}(j,\tau)|^2U_{A22}^*(j-
1,\tau)+U_{A11}(0,\tau)|U_{B11}(0,\tau)|^2]+\,\nonumber\\
&&+A_1^*A_2|B_1|^2\sum_{j=0}^{\infty}|f_{j+1}(x)|^2U_{A11}(j+1,\tau)|U_{B12}(j,
\tau)|^2U_{A22}^*(j,\tau)\}
\end{eqnarray}
\begin{eqnarray}
\label{RHOA14APP_A}&&\rho_{14}(x)=|c_0(x)|^2
\{|A_2|^2|B_2|^2\sum_{j=1}^{\infty}f_{j}(x)f_{j-1}^*(x)
U_{A11}(j,\tau)U_{B11}(j,\tau)U_{A21}^*(j-1,\tau)U_{B21}^*(j-
1,\tau)+\,\nonumber\\
&&+A_1^*A_2B_1^*B_2[\sum_{j=1}^{\infty}
|f_{j}(x)|^2U_{A11}(j,\tau)U_{B11}(j,\tau)U_{A22}^*(j-1,\tau)U_{B22}^*(j-
1,\tau)+U_{A11}(0,\tau)U_{B11}(0,\tau)]+\,\nonumber\\
&&+|A_2|^2|B_1|^2[\sum_{j=1}^{\infty}f_{j}(x)f_{j-1}^*(x)U_{A11}(j+1,\tau)
U_{B12}(j,\tau)U_{A21}^*(j,\tau)U_{B22}^*(j-1,\tau)+ \,\nonumber\\
&&+f_1(x)U_{A11}(1,\tau)U_{B12}(0,\tau)U_{A21}^*(0,\tau)]+\,\nonumber\\
&&+|A_1|^2|B_2|^2[\sum_{j=1}^{\infty}f_{j+1}(x)f_{j}^*(x)U_{A12}(j,\tau)
U_{B11}(j+1,\tau)U_{A22}^*(j-1,\tau)U_{B21}^*(j,\tau)+\,\nonumber\\
&&+ f_1(x)U_{A12}(0,\tau)U_{B11}(1,\tau)U_{B12}^*(0,\tau)]+\,\nonumber\\\
&&+|A_1|^2|B_1|^2[\sum_{j=1}^{\infty}f_{j+1}(x)f_{j}^*(x)U_{A12}(j,\tau)
U_{B12}(j,\tau)U_{A22}^*(j-1,\tau)U_{B22}^*(j-1,\tau)+\,\nonumber\\
&&+f_1(x)U_{A12}(0,\tau) U_{B12}(0,\tau)]+\,\nonumber\\
&&+A_1A_2^*B_1B_2^*\sum_{j=1}^{\infty} f_{j+1}(x)f_{j-
1}^*(x)U_{A12}(j,\tau)U_{B12}(j,\tau) U_{A21}^*(j-1,\tau)U_{B21}^*(j-1,\tau)\}
\end{eqnarray}
\begin{eqnarray}
\label{RHOA23APP_A}&&\rho_{23}(x)=|c_0(x)|^2
A_1^*A_2B_1B_2^*[\sum_{j=0}^{\infty}|f_{j}(x)|^2U_{A11}(j,\tau)U_{B22}(j-
1,\tau)U_{A22}^*(j-
1,\tau)U_{B11}^*(j,\tau)+U_{A11}(0,\tau)U_{B11}^*(0,\tau)]\,\nonumber\\
\end{eqnarray}
\begin{eqnarray}
\label{RHOA24APP_A} &&\rho_{24}(x)=|c_0(x)|^2
\{|A_2|^2B_1B_2^*\sum_{j=1}^{\infty}f_{j}(x)f_{j-
1}^*(x)U_{A11}(j,\tau)U_{B22}(j-1,\tau)U_{A21}^*(j-1,\tau)U_{B21}^*(j-
1,\tau)+\,\nonumber\\
&&+|A_1|^2B_1B_2^*[\sum_{j=1}^{\infty}f_{j+1}(x)f_{j}^*(x)U_{A12}(j,\tau)U_{B22}
(j,\tau)U_{A22}^*(j-1,\tau)U_{B21}^*(j,\tau)+\,\nonumber\\
&&+f_1(x)U_{A12}^(0,\tau)U_{B22}(0,\tau)U_{B21}^*(0,\tau)]+\,\nonumber\\
&&+A_1^*A_2|B_2|^2[\sum_{j=1}^{\infty}|f_{j}(x)|^2
U_{A11}(j,\tau)|U_{B21}(j,\tau)|^2U_{A22}^*(j-
1,\tau)+U_{A11}(0,\tau)|U_{B21}(0,\tau)|^2]+\,\nonumber\\
&&+A_1^*A_2|B_1|^2[\sum_{j=1}^{\infty}|f_{j}(x)|^2U_{A11}(j,\tau)|U_{B22}(j-
1,\tau)|^2U_{A12}^*(j-1,\tau)+U_{A11}(0,\tau)]\}
\end{eqnarray}
\begin{eqnarray}
\label{RHOA34APP_A}  &&\rho_{34}(x)=|c_0(x)|^2
\{|A_2|^2B_1^*B_2[\sum_{j=1}^{\infty}|f_{j}(x)|^2|U_{A21}(j,\tau)|^2U_{B11}(j,
\tau)U_{B22}^*(j-1,\tau)+|U_{A21}(0,\tau)|^2U_{B11}(0,\tau)\,\nonumber\\
&+&|A_1|^2B_1^*B_2[\sum_{j=1}^{\infty}|f_{j}(x)|^2U_{A22}(j-
1,\tau)U_{B11}(j,\tau)U_{A22}^*(j-1,\tau)U_{B22}^*(j-
1,\tau)+U_{B11}(0,\tau)]\,\nonumber\\
&+&A_1A_2^*|B_2|^2\sum_{j=1}^{\infty}f_{j}(x)f_{j-1}^*(x) U_{A22}(j-
1,\tau)U_{B11}(j,\tau)U_{A21}^*(j-1,\tau)U_{B21}^*(j-1,\tau)\,\nonumber\\
&+&A_1^*A_2|B_1|^2[\sum_{j=1}^{\infty}f_{j+1}(x)f_{j}^*(x)U_{A22}(j,\tau)U_{B12}
(j,\tau)U_{A21}^*(j,\tau)U_{B22}^*(j-
1,\tau)+f_1(x)U_{A22}(0,\tau)U_{B12}(0,\tau)U_{A21}^*(0,\tau)]\}\,\nonumber\\
\end{eqnarray}
\end{widetext}

\end{document}